\documentclass[%
reprint,nofootinbib,
superscriptaddress,
 amsmath,amssymb,
 aps,
]{revtex4-2}

\usepackage{graphicx}% Include figure files
\usepackage{dcolumn}% Align table columns on decimal point
\usepackage{bm}% bold math
\usepackage{caption}
\usepackage{subcaption}
\DeclareCaptionJustification{justified}{\justifying}
\usepackage{slashed}
\usepackage[utf8]{inputenc}
\usepackage{braket}
\usepackage{ragged2e}

\usepackage[compat=1.1.0]{tikz-feynman}

\newcommand{\bspinor}{\overline{u}_{s^{\prime}}(p^\prime)}
\newcommand{\spinor}{u_s(p)}
\newcommand{\bspshort}{\overline{u}_{s^{\prime}}}
\newcommand{\spshort}{u_s} 
\newcommand{\qedvert}{\left[\Gamma^{A}\right]^{\mu}}
\newcommand{\Gamfun}[1]{ g^2 \int \dfrac{d^Dl}{i(2\pi)^D} \dfrac{ #1  }{l^2 D_{p^\prime+l} D_{p+l} } }
\newcommand{\Gamfuns}{ g^2 \int \dfrac{d^Dl}{i(2\pi)^D} \dfrac{ 1  }{l^2} }
\newcommand{\GamfunNA}[1]{ g^2 \int \dfrac{d^Dl}{i(2\pi)^D} \dfrac{ #1  }{ q^{\prime 2} q^2 D_l } }

\begin{document}

\begin{abstract}
    
\end{abstract}

\title{One-loop off-shell quark-gluon vertex  in arbitrary gauge and dimensions: \\a streamlined approach through the second-order formalism of QCD} 

\author{Victor Miguel Banda Guzm\'an}
\email{victor.banda@upslp.edu.mx}
\affiliation{Universidad Polit\'ecnica de San Luis Potos\'i, Urbano Villal\'on 500, Col. La Ladrillera, C.P. 78363, San Luis Potos\'i, S.L.P., M\'exico.}
\author{A. Bashir}
\email{adnan.bashir@umich.mx; adnan.bashir@dci.uhu.es}
\affiliation{Instituto de F\'isica y Matem\'aticas, Universidad
Michoacana de San Nicol\'as de Hidalgo, Edificio C-3, Ciudad
Universitaria, Morelia, Michoac\'an 58040, M\'exico}
\affiliation{Department of Integrated Sciences and Center for Advanced Studies in Physics, Mathematics and Computation, University of Huelva, E-21071 Huelva, Spain}

\begin{abstract}

The standard Feynman rules used for perturbative calculations in quantum chromodynamics (QCD) are derived from a Lagrangian that is first-order in derivatives. It includes a three-point quark-gluon vertex which obscures the precise disentangled manner in which spin and momentum are interchanged during these interactions. An unambiguous understanding of this interchange is insightful for efficiently extracting physically relevant information from various Green's functions. 
To separate the scalar and spin degrees of freedom and gain physical insight from the outset, we examine the quark-gluon vertex using the less commonly employed second-order formalism of QCD. We compute this off-shell vertex in arbitrary space-time dimensions and covariant gauges by using scalar integrals with shifted dimensions, which include higher powers of the propagators, within a combined first- and second-order formalism. This approach naturally identifies the transverse components of the quark-gluon vertex, even before evaluating the tensor Feynman integrals. Our results are in complete agreement with those obtained from the first-order formalism.
We also compute the on-shell version of this vertex using exclusively the second-order formalism, 
facilitating a precise identification of spin and momentum interchange. Through analyzing the Pauli form factor at $k^2=0$ (where $k$ represents the momentum of the external gluon), we find that only a specific set of second-order Feynman diagrams
%—corresponding to the type of interaction involved
are relevant for calculating the electromagnetic and chromomagnetic dipole moments. These diagrams represent quantum processes in which the spin of the incoming quark changes only once due to interactions with the virtual gluons that form the quark-gluon vertex. All other interactions involve only momentum interchange (scalar interactions). Furthermore, we confirm existing results in the literature, which suggest that defining the chromomagnetic dipole moment in the limit $k^2 \rightarrow 0$ leads to infrared divergences. 
%Therefore, if we insist on an acceptable definition of the chromomagnetic dipole moment within perturbation theory, we might seek an alternative kinematic configuration which should yield a finite and gauge independent result for this {\em observable}.       

\end{abstract}

\maketitle

\section{Introduction}

\begin{figure}[b!]
\begin{subfigure}[b]{0.45\textwidth}
\includegraphics[scale=0.5]{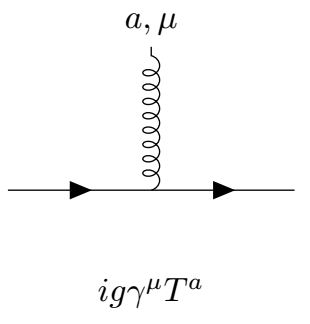}
\caption{First-order three-point quark-gluon vertex.} \label{FO_3p_qg_v}
\end{subfigure}
\begin{subfigure}[b]{0.45\textwidth}
\includegraphics[scale=0.35]{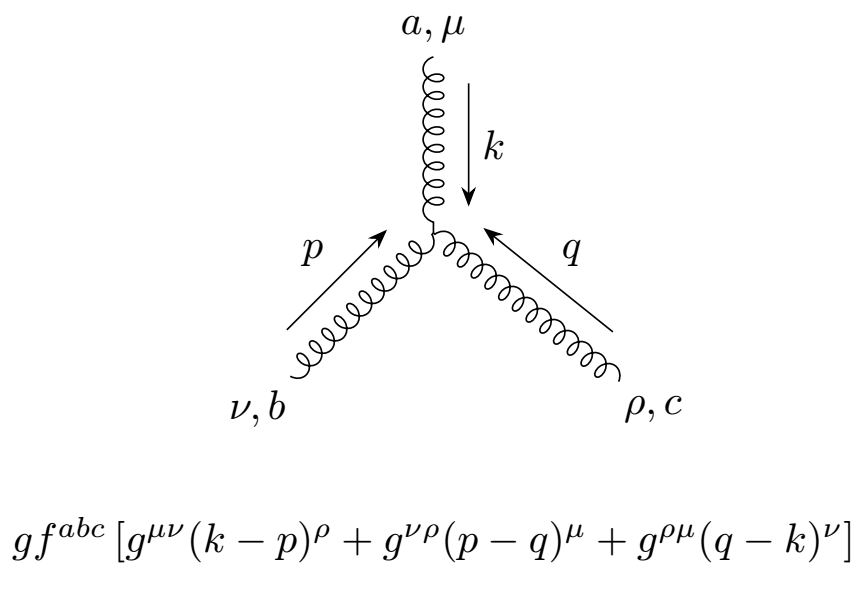}
\caption{Three-gluon vertex.} \label{3g_vert}
\end{subfigure}
\begin{subfigure}[b]{0.45\textwidth}
\includegraphics[scale=0.35]{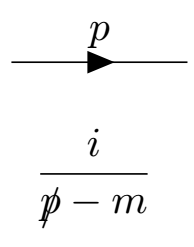}
\caption{Fermion propagator.} \label{f_prop}
\end{subfigure}
\begin{subfigure}[b]{0.45\textwidth}
\includegraphics[scale=0.35]{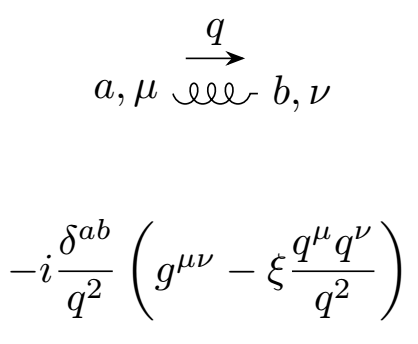}
\caption{Gluon propagator in arbitrary covariant gauge.} \label{g_prop}
\end{subfigure}
\caption{First-order Feynman rules for fermion and gluon vertices and propagators.}
\label{FO_rules}
\end{figure}

The conventional approach to perturbative calculations in quantum chromodynamics (QCD) relies on Feynman diagrams, with the rules illustrated in Fig.~\ref{FO_rules}.
These rules are derived from the following Lagrangian,
\begin{eqnarray}
    \mathcal{L} = - \dfrac{1}{4} \left(F^{a}_{\mu\nu}\right)^2 + \overline{\psi} \left(i\slashed{D}-m\right) \psi - \frac{1}{2\xi} (\partial_{\mu} A^{\mu})^2 \,.
\end{eqnarray}
This Lagrangian is first-order in the derivatives. The field strength tensor $F_{\mu\nu}^a$ reads as,
\begin{eqnarray}
    F_{\mu\nu}^a = \partial_\mu A_\nu^a - \partial_\nu A^a_\mu + g f^{abc} A_\mu^b A_\nu^c \,.
\end{eqnarray}
Here $f^{abc}$ are the structure constants of the $SU(N)$ group, and the covariant derivative $D_\mu$ is defined as
\begin{eqnarray}
    D_\mu = \partial_\mu - i g A^a_\mu T^a \,,
\end{eqnarray}
where $T^a$ are the generators of the $SU(N)$ group. 

Although the Feynman rules illustrated in Fig.~\ref{FO_rules} can be implemented straightforwardly, they often lead to cumbersome and lengthy calculations that blur the underlying physical information of a quantum field function. 
For instance, in the context of quantum electrodynamics (QED), the anomalous magnetic dipole moment of the electron at the one-loop level is determined to be equal to ${\alpha}/{2\pi}$, where $\alpha={e^2}/{4\pi}$ is the fine-structure constant. The higher order contributions to this physical observable emerge following extensive calculations of the electromagnetic vertex at the multi-loop level. It is desirable and perhaps conceivable that the final simple results could be derived through a more direct approach. It may be tempting to develop an alternative mathematical framework and computational strategy for perturbative calculations in both QCD and QED, one that enhances the extraction of physical insight from the outset and makes the computation of physical observables more manifest. The second-order formalism of QCD developed in Refs.~\cite{Hostler,Morgan} may serve as a valuable tool in developing such a worthwhile framework. By decoupling the scalar and spin degrees of freedom in the interactions between quarks and gluons, this formalism can provide a transparent picture of how quarks and gluons exchange spin and momentum. This methodology provides an alternative but equivalent approach to the first-order standard formalism. It has lately resurfaced in the literature as providing a connection between the worldline formalism of quantum field theory and the approach of Feynman diagrams, for details see Refs.~\cite{Schubert_WL, Worldline_openflines}. It has also been applied recently to the study of the electromagnetic one-loop vertex in standard and reduced QED, Refs.~\cite{QEDVer,RQED}. 

A simple example of how the second-order formalism decouples the scalar and spin degrees of freedom in the interaction of the quarks and gluons is given by the on-shell quark-gluon vertex $\left[\Gamma^{\text{OS}}_{\text{tree}}\right]^{a\mu}$ at tree level. In the first-order conventional approach, the vertex is given by the Fig.~\ref{FO_3p_qg_v}, while in the second-order formalism it directly rewrites as (see Figs.~\ref{SecondOR_3pointvert}),
\begin{eqnarray}
    \left[\Gamma^{\text{OS}}_{\text{tree}}\right]^{a\mu} = \dfrac{igT^a}{2m}\left[ (p+p^\prime)^\mu - \sigma^{\mu\alpha} k_\alpha \right] \,, \label{onshell_vertex_tree}
\end{eqnarray}
where $k$ corresponds to the gluon momentum, while $p$ and $p^\prime$ are the momenta of the incoming and outgoing quarks, respectively. Eq.~\eqref{onshell_vertex_tree} depicts a clearcut separation between scalar and spin degrees of freedom in the quark-gluon interaction. Additionally, there is a natural decomposition between longitudinal and transverse components with respect to the gluon momenta $k$. This separation gets entangled and obscure in the first order formalism. 

\begin{figure}[h!]
\begin{subfigure}[b]{0.45\textwidth}
\includegraphics[scale=0.40]{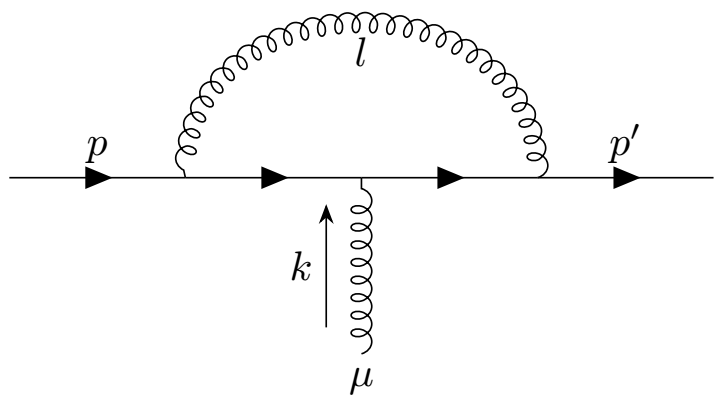}
\caption{Abelian contribution to the quark-gluon vertex.} \label{Abelian_graph}
\end{subfigure}
\begin{subfigure}[b]{0.45\textwidth}
\includegraphics[scale=0.40]{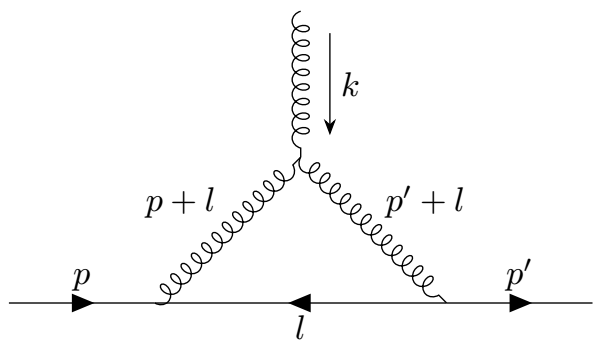}
\caption{Non-abelian contribution to the quark-gluon vertex.} \label{NonAbelian_graph}
\end{subfigure}
\caption{Feynman diagrams for the one-loop quark-gluon vertex within the first-order formalism.}
\label{FO_diagrmas_qgv}
\end{figure}

The quark-gluon vertex is a fundamental component of QCD which has extensively been studied both perturbatively and non-perturbatively~\cite{Davydychev,Fischer:2003rp,Alkofer:2008tt,Williams:2014iea,Williams:2015cvx,Bermudez:2017bpx,Sultan:2018tet,Albino:2021rvj,Lessa:2022wqc}. In this work, we employ the second-order formalism to examine the one-loop quark-gluon vertex both in the on-shell and off-shell configurations in arbitrary space-time dimensions and covariant gauges. The relevant Feynman diagrams for the one-loop calculations of the quark-gluon vertex using the first-order approach are presented in Fig.~\ref{FO_diagrmas_qgv}. The diagram in Fig.~\ref{Abelian_graph} describes the abelian contribution to the vertex, which can also be used to derive the one-loop electromagnetic vertex.  Fig.~\ref{NonAbelian_graph} depicts the non-abelian contribution which is a characteristic of QCD. Employing the second-order formalism, we derive both longitudinal and transverse components of the vertex distinctly from the outset. 
%In the on-shell case, this approach enables a detailed analysis of the interactions between the one-loop virtual quarks and gluons associated with the vertex.

The article is organized as follows: Sec.~\ref{preeliminaries} contains an introduction to the second-order formalism, and how it can be obtained from the first-order formalism. This section also contains a review of the main tools and methods for evaluating one-loop three-point Feynman tensor and scalar integrals that are used in the rest of the article. In Sec.~\ref{Offshell section}, we introduce the second-order formalism for the evaluation of the off-shell one loop quark-gluon vertex in arbitrary dimensions and covariant gauge. A combination of first- and second-order approach is more suitable for this computation. This course of action allows us to separate out the transverse components of the quark-gluon vertex
in a natural manner from the onset
when the external gluon is attached to the quark line. In Sec.~\ref{Onshell_section}, we perform the strategy outlined in Ref.~\cite{Morgan} to make a transition from the first-order formalism to the second-order formalism completely. Here, we use the Feynman gauge to simplify the discussion. We present the different contributions from the scalar and spin interaction between quarks and gluons to the Pauli form factor at $k^2=0$, and thus to the electromagnetic and chromomagnetic dipole moments. It also contains the Feynman diagrams that arise from the second-order formalism for the one-loop quark-gluon vertex.  
The concluding remarks are provided in Sec.~\ref{Conclusions}. The manuscript is complemented with three appendices which contain supplemental information. Appendix~\ref{coefficients} contains expressions with scalar integrals for the non abelian contributions to the off-shell quark-gluon vertex. In Appendix~\ref{SO_vert_fun} we provide the on-shell contributions to the quark-gluon vertex in arbitrary dimensions that arise from the second-order formalism. For the evaluation of the dipole moments in $D=4$ dimensions, we provide the evaluation of the main on-shell scalar integrals with higher power in the propagators and shifted dimensions at $k^2=0$ in Appendix~\ref{scalar_int_onshell}. We use the dimensional regularization scheme to regularize ultraviolet divergences. 

\section{Preliminaries} \label{preeliminaries}

In this section, we review the main features of the second-order formalism for QCD, and the techniques that we  employ in the evaluation of the Feynman integrals that show up in the calculations. 

\subsection{Second-order formalism for QCD}

\begin{figure}[h!]
\includegraphics[scale=0.5]{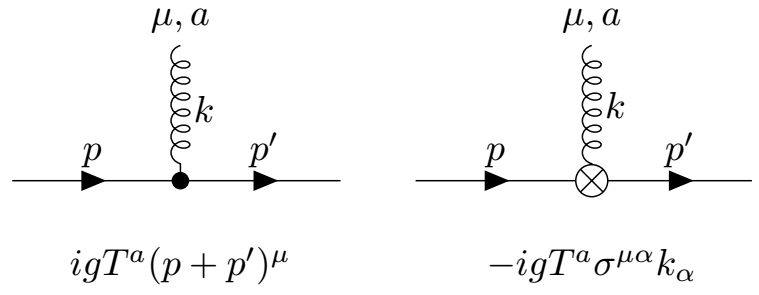}
\caption{Three-point vertex of the second-order formalism for QCD, separated into its scalar (left) and spin (right) contributions.} \label{SecondOR_3pointvert}
\end{figure}

The second-order formalism for fermions in gauge theories was first derived in Refs.~\cite{Hostler, Morgan}. It disentangles the gauge interaction of the fermionic particle into scalar and spin degrees of freedom with
appropriate usage of Gordon-like identities. It can be achieved, see~\cite{Morgan}, by formally rewriting the product of the fermion propagator $S(p+k)$ with the first-order vertex $i g T^a \gamma^\mu$ as,
\begin{eqnarray}
    S(p+k) (i g T^a \gamma^\alpha) &=& \left(i \dfrac{\slashed{p}+\slashed{k}+m}{(p+k)^2-m^2}\right)\left( i g T^a \gamma^\alpha \right) \,, \nonumber \\
    &=& - g T^a \dfrac{A_{p,k}^\alpha}{D_{p+k}} \,, \label{PropVertFO}
\end{eqnarray}
where
\begin{align}
    A_{p,k}^\alpha &= B^{\alpha}_{p,k} - C_p^{\alpha} \,, & B_{p,k}^{\alpha} &= 2p^\alpha + k^\alpha - \sigma^{\alpha\beta}k_\beta \,, \nonumber \\
    C_p^\alpha &= \gamma^\alpha(\slashed{p}-m) \,, & D_q &= q^2-m^2 \,. \label{Def_SO} 
\end{align}
Here, $\sigma^{\mu\nu}=[\gamma^\mu,\gamma^\nu]/2$, $i D^{-1}_q$ represents the scalar propagator, and $B_{p,k}$ corresponds to the total colorless three-point vertex of the second-order formalism (see Fig.~\ref{SecondOR_3pointvert}), which separates the scalar and spin gauge interactions. Thus a product of a propagator and a first-order vertex decomposes into a second-order contribution and a leftover first-order term given by the operator $C_p^\alpha$.

\begin{figure}[h!]
\includegraphics[scale=0.4]{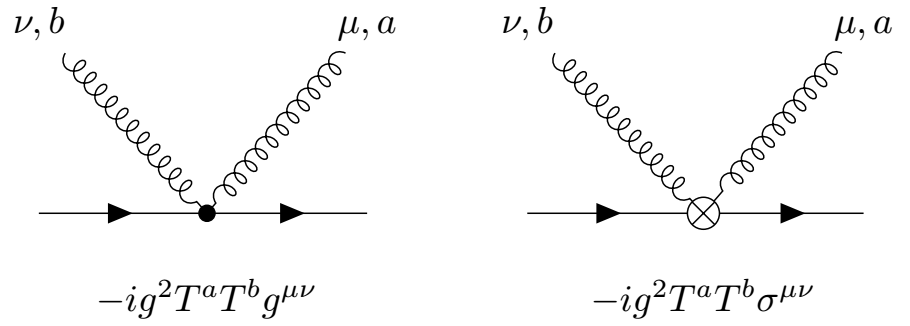}
\caption{Four-point vertex of the second-order formalism for QCD, separated into its scalar (left) and spin (right) contributions.} \label{SecondOR_4pointvert}
\end{figure}

\begin{figure}[h!]
\includegraphics[scale=0.5]{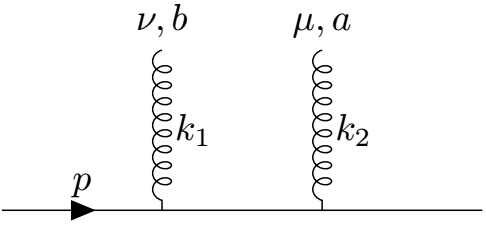}
\caption{ Subdiagram that produces the contribution to the four-point vertex in the second-order formalism.} \label{Subdiagram_4pointvert}
\end{figure}

Additionally to the three-point vertex, in the second-order formalism there is also a four-point vertex whose scalar and spin contributions are displayed in Fig.~\ref{SecondOR_4pointvert}. It arises when there are two consecutive pairs of a fermion propagator and a first-order vertex in a Feynman diagram, see Fig.~\ref{Subdiagram_4pointvert}. An amplitude
constructed from a first-order Feynman diagram that contains
a subdiagram as the one shown in Fig.~\ref{Subdiagram_4pointvert} is proportional to the operator $M^{\mu\nu, ab}(p,k_1,k_2)$ which is given by,
\begin{eqnarray}
    M^{\mu\nu, ab}(p,k_1,k_2) && \nonumber \\
    && \hspace{-50pt} = -g^2 T^a T^b S(p+k_1+k_2) \gamma^\mu S(p+k_1) \gamma^\nu \,.
\end{eqnarray}
According to Eq.~\eqref{PropVertFO}, this operator can be rewritten as
\begin{eqnarray}
    M^{\mu\nu, ab}(p,k_1,k_2) &=& g^2 T^a T^b \dfrac{A^\mu_{p+k_1,k_2}A_{p+k_1}^\nu}{D_{p+k_1+k_2}D_{p+k_1}} \,. \label{M_operator_A}
\end{eqnarray}
Since
\begin{eqnarray}
    C_{p+k}^\alpha A_{p,k}^\beta &=& D_{p+k} \gamma^\alpha \gamma^\beta \,, \nonumber \\  
    &=& D_{p+k} ( \sigma^{\alpha \beta} + g^{\alpha\beta} ) \,, \label{CAIdentity}
\end{eqnarray}
Eq.~\eqref{M_operator_A} acquires the following form:
\begin{eqnarray}
    M^{\mu\nu, ab}(p,k_1,k_2) &=& g^2 T^a T^b \left[ \dfrac{B^\mu_{p+k_1,k_2}A^\nu_{p+k_1}}{D_{p+k_1+k_2}D_{p+k_1}} \right. \nonumber \\
    && \left. \hspace{1.5cm}- \dfrac{g^{\mu\nu}+\sigma^{\mu\nu}}{D_{p+k_1+k_2}} \right] \,. 
\end{eqnarray}
Thus, the multiplication of fermion propagators and first-order vertices generates a term with one less propagator. It readily gets identified as a four-point vertex.  

The details of the transition between the first and second-order formalism can be found in Ref.~\cite{Morgan}.

\vspace{-.5cm}

\subsection{Summary of useful computational techniques}

In the evaluation of the quark-gluon vertex at one-loop we need to perform tensor Feynman integrals with the following structure:
\begin{eqnarray}
    J^{\mu}(a_1,a_2,a_3) &=& \int \dfrac{d^Dl}{i \pi^{\frac{D}{2}}} \dfrac{l^\mu}{ \prod\limits_{i=1}^3 D_i^{a_i} } \,, \nonumber \\
    J^{\mu\nu}(a_1,a_2,a_3) &=& \int \dfrac{d^Dl}{i \pi^{\frac{D}{2}}} \dfrac{l^\mu l^\nu}{ \prod\limits_{i=1}^3 D_i^{a_i} } \,, \label{Def_TensorInt}
\end{eqnarray}
where the scalar propagators $D_i$ are given by
\begin{eqnarray}
    D_i = (p_i+l)^2-m_i^2 \,.
\end{eqnarray}
One way to achieve this goal is to employ the tensor reduction algorithm described in Refs.~\cite{Anastasiou1, Anastasiou2, Fiorentin, TarasovTR}. This method uses the Schwinger parametrization of Feynman integrals to obtain identities between tensor and scalar Feynman integrals.

In our case, the one-loop scalar Feynman integral with three propagators $J^D_3(a_1,a_2,a_3)$, defined as 
\begin{eqnarray}
    J^D_3(a_1,a_2,a_3) = \int \dfrac{d^Dl}{i \pi^{\frac{D}{2}}} \dfrac{1}{\prod\limits_{i=1}^3 D_i^{a_i}} \,, \label{def_J3ScalarIntegral}
\end{eqnarray}
in the Schwinger parametrization reads as
\begin{eqnarray}
    J^D_3(a_1,a_2,a_3) = \int \mathcal{D}x \dfrac{1}{c^{D/2}} \text{e}^{\frac{P}{c}-f} \,, \label{J3_SP}
\end{eqnarray}
where 
\begin{eqnarray}
    \int \mathcal{D}x &=& \prod_{i=1}^3 \dfrac{(-1)^{a_i}}{\Gamma(a_i)} \int_0^\infty x_i^{a_i-1}\, dx_i \,, \nonumber \\
    c &=& \sum_{i=1}^3 x_i \,, \qquad f=\sum_{i=1}^3x_im_i^2 \,, \nonumber \\
    P &=& - \sum_{i,j=1}^3 x_i x_j p_i \cdot p_j + c \sum_{i=1}^3 x_i p_i^2 \,.
\end{eqnarray}
Here, the subscript in $J^D_3(a_1,a_2,a_3)$ represents the number of massive propagators, just as in Ref.~\cite{Davydychev}, while the superscript represents the number of space-time dimensions. 

The tensor integral $J^{\mu}(a_1,a_2,a_3)$ defined in Eq.~\eqref{Def_TensorInt} in the Schwinger representation is given by,
\begin{eqnarray}
    J^{\mu}(a_1,a_2,a_3) = - \int \mathcal{D}x \dfrac{\sum_{i=1}^3 x_i p_i^\mu}{c^{(D+2)/2 }} \text{e}^{\frac{P}{c}-f} \,. \label{Jmu_3_SP}
\end{eqnarray}
Comparing Eq.~\eqref{Jmu_3_SP} with Eq.~\eqref{J3_SP}, we obtain,
\begin{eqnarray}
    J^{\mu}(a_1,a_2,a_3) = \sum_{i=1}^3 a_i p_i^\mu \, \mathbf{i}^+ J_3^{D+2}(a_1,a_2,a_3) \,, \label{TI_1}
\end{eqnarray}
where the operator $\mathbf{i}^+$ increases the power of the i-th propagator:
\begin{eqnarray}
    \mathbf{i}^+ J_3^{D+2}(...,a_i,...) = J_3^{D+2}(...,a_i+1,...) \,.
\end{eqnarray}
Similarly, one can readily obtain the following identity for $J^{\mu\nu}(a_1,a_2,a_3)$,
\begin{eqnarray}
    && J^{\mu\nu}(a_1,a_2,a_3) = -\dfrac{1}{2} g^{\mu\nu} J^{D+2}_3(a_1,a_2,a_3) \nonumber \\
    && \hspace{10pt} + \sum_{i,j=1}^3 c_{ij} p_i^\mu p_j^\nu \mathbf{i}^+ \mathbf{j}^+ J^{D+4}_3(a_1,a_2,a_3) \,, \label{TI_2}
\end{eqnarray}
where the coefficients $c_{ij}$ read as
\begin{eqnarray}
    c_{ij} = \begin{cases}
        a_i(a_i+1) & \text{for } i=j\\
        a_i a_j & \text{for } i \neq j \,.
    \end{cases}
\end{eqnarray}
Eqs.~\eqref{TI_1} and~\eqref{TI_2} are particular examples of the general formulae obtained in Ref.~\cite{DavydychevOneLoopInt}. Using these identities, we can transform the tensor integrals of the one-loop quark-gluon vertex into scalar integrals with shifted space-time dimensions and higher powers in the propagators. 

To reduce the dimension of the scalar integrals, we can use the method introduced in Ref.~\cite{DimensionalReduction}. According to this method, we again use the Schwinger representation of the scalar integral $J_3^D(a_1,a_2,a_3)$ in Eq.~\eqref{J3_SP} to obtain the following identity,
\begin{eqnarray}
    \sum_{i=1}^3 \dfrac{\partial}{\partial m_i^2} J^D_3(a_1,a_2,a_3) = - \int \mathcal{D}x \dfrac{\sum_{i=1}^3x_i}{c^{D/2}} \text{e}^{\frac{P}{c}-f} \,.
\end{eqnarray}
Since $\sum\limits_{i=1}^3 x_i = c$, from the equation above, we obtain
\begin{eqnarray}
    J^{D-2}_3(a_1,a_2,a_3) = - \sum_{i=1}^3 a_i\, \mathbf{i}^+ J^{D}_3(a_1,a_2,a_3) \,. \label{DimensionalReduction1}
\end{eqnarray}
Now, for the computation of the one-loop quark-gluon vertex, we need the following set of scalar integrals:
\begin{eqnarray}
    J_1^D(a_1,a_2,a_3) &=& \int \dfrac{d^Dl}{i \pi^{\frac{D}{2}}} \dfrac{1}{(p^\prime+l)^{2a_1}(p+l)^{2a_2}(l^2-m^2)^{a_3}}, \nonumber \\
    J_2^D(a_1,a_2,a_3) &=& \int \dfrac{d^Dl}{i \pi^{\frac{D}{2}}} \dfrac{1}{D_{p^\prime+l}^{a_1} D_{p+l}^{a_2} (l^2)^{a_3}} \,,  \label{scIntqg}
\end{eqnarray}
which correspond to $J_3^D(a_1,a_2,a_3)$ defined in Eq.~\eqref{def_J3ScalarIntegral} with $m_1=m_2=0$, $m_3=m$ for $J_1^D(a_1,a_2,a_3)$, and $m_1=m_2=m$, $m_3=0$ for $J_2^D(a_1,a_2,a_3)$, while fixing the momentum $p_1=p^\prime$, $p_2=p$ and $p_3=0$ in both scalar integrals.  

To obtain more useful identities similar to Eq.~\eqref{DimensionalReduction1} for reducing the dimension of the scalar integrals, we continue with the approach outlined in 
Ref.~\cite{DimensionalReduction}. There, the following identity is used:
\begin{eqnarray}
  \hspace{-4mm}  \int \dfrac{d^Dl}{i \pi^{\frac{D}{2}}} \dfrac{\partial}{\partial_\mu}\left[ \left( x_3^{(i)}l_\mu + x_1^{(i)} p^\prime_\mu + x_2^{(i)} p_\mu \right) A^{(i)}\right] = 0 \,, \label{DimRedIden}
\end{eqnarray}
where, in our case, we define the functions $A^{(i)}$, with $i=1,2$, as, 
\begin{eqnarray}
    A^{(1)} &=& \dfrac{1}{(p^\prime+l)^{2a_1}(p+l)^{2a_2}(l^2-m^2)^{a_3}} \,, \nonumber \\
    A^{(2)} &=& \dfrac{1}{D_{p^\prime+l}^{a_1} D_{p+l}^{a_2} (l^2)^{a_3}} \,. 
\end{eqnarray}
The identity in Eq.~\eqref{DimRedIden} is valid for any $x_i^{(1)}$. But, if we fixed the values of the $x_i^{(1)}$ such that
\begin{eqnarray}
    x_1^{(1)} &=& \dfrac{2}{Y^{(1)}}\left( p^2 p^\prime \cdot k - m^2 p \cdot k \right) \,, \nonumber \\
    x_2^{(1)} &=& -\dfrac{2}{Y^{(1)}}\left( p^{\prime 2} p \cdot k - m^2 p^\prime \cdot k \right) \,, \nonumber \\
    x_1^{(2)} &=& \dfrac{2}{Y^{(2)}}\left( p^2 p^\prime \cdot k + m^2 p \cdot k \right) \,, \nonumber \\
    x_2^{(2)} &=& -\dfrac{2}{Y^{(2)}}\left( p^{\prime 2} p \cdot k + m^2 p^\prime \cdot k \right) \,, \nonumber \\
    x_3^{(i)} &=& \dfrac{4G}{Y^{(i)}} \,, \quad i=1,2,
\end{eqnarray}
where $k=p^\prime-p$, $G$ is the Gram determinant
\begin{eqnarray}
    G = p^2 p^{\prime 2} - (p \cdot p^\prime)^2 ,,
\end{eqnarray}
and $Y^{(i)}$ are the determinant of the modified Caley matrix for the scalar integrals $J_i^D$:
\begin{eqnarray}
    Y^{(1)} &=& - 2 k^2( m^4 - 2m^2 p \cdot p^\prime + p^2 p^{\prime 2} ) \,, \nonumber \\
    Y^{(2)} &=& -2 \left[ k^2(m^4+p^2p^{\prime 2})+2m^2( p^{\prime 2} p \cdot p^\prime \right. \nonumber \\
    &&  \hspace{8mm} \left. + p^2(p \cdot p^\prime-2p^{\prime 2}) ) \right] \,.
\end{eqnarray}
From Eqs.~\eqref{DimensionalReduction1} and ~\eqref{DimRedIden}, we obtain 
\begin{eqnarray}
    J_i^{D+2}(a_1,a_2,a_3) &=& - \dfrac{1}{x_3^{(i)}(d+1-\sum\limits_{j=1}^3a_j)}\left[ J_i^D(a_1,a_2,a_3) \right. \nonumber \\
    && \hspace{-70pt} + x_1^{(i)} J_i^D(a_1-1,a_2,a_3) + x_2^{(i)} J_i^{D}(a_1,a_2-1,a_3) \nonumber \\
    && \left. \hspace{-70pt} + ( x_3^{(i)}-x_1^{(i)}-x_2^{(i)} ) J_i^D(a_1,a_2,a_3-1) \right] \,, \label{DimRedJi}
\end{eqnarray}
for $i=1,2$. The generalization of this identity to $n$-points with arbitrary masses can be found in Ref.~\cite{DimensionalReduction}. 

Now that we have the tensor integrals identities (Eqs.~\eqref{TI_1} and \eqref{TI_2}), and the formula to reduce the dimension of scalar integrals (Eq.~\eqref{DimRedJi}), we can use, if needed, the well-known method of integration by parts, Refs.~\cite{IBP_alg, IBP_letter}, to rewrite all scalar integrals into the sets $S_1$ and $S_2$ of master integrals: 
\begin{eqnarray}
    S_1 &= \{& J_1^D(0,0,1), J_1^D(0,1,1), J_1^D(1,0,1), \nonumber \\ 
    && J_1^D(1,1,0), J_1^D(1,1,1) \}, \nonumber \\
    S_2 &= \{& J_2^D(0,1,0), J_2^D(0,1,1), J_2^D(1,0,1), \nonumber \\
    && J_2^D(1,1,0), J_2^D(1,1,1)  \}. \label{MasterInt}
\end{eqnarray}
For this purpose we can use, for example, the symbolic programming package LiteRed~\cite{LiteRed1, LiteRed2}. 

\section{ Off-shell quark-gluon vertex } \label{Offshell section}

In the case of an off-shell quark-gluon vertex, we can not completely introduce the second-order formalism since there are not enough pairs of propagators and the first-order vertex to rewrite them in terms of scalar propagators and second-order vertices. Thus, for the off-shell case, we combine the first and second-order formalisms by rewriting all possible pairs of a fermion propagators and first-order vertex using Eq.~\eqref{PropVertFO}. This procedure leads to a natural decomposition between the transverse and longitudinal components with respect to the  external gluon momentum, when this external momentum is attached to a fermion line as in the case of the abelian graph (Fig.~\ref{Abelian_graph}). 

We first present a summary of the combined first- and second-order formalism applied to the abelian graph (further details can be found in Ref.~\cite{QEDVer}). Then, in the preceding subsection we apply this method to the non-abelian graph (Fig.~\ref{NonAbelian_graph}).

\begin{widetext}
\subsection{Off-shell abelian graph}

Fig.~\ref{Abelian_graph} depicts the abelian Feynman diagram for the vertex function $\Gamma^{\mu a}$ at one-loop within the first-order formalism. It is basically the QED vertex which has also been extensively studied in the literature~\cite{Curtis:1990zs,Bashir:1999bd,Bashir:2000rv,Bashir:2007qq,Kizilersu:2009kg,Albino:2018ncl}.
Denoting this abelian contribution as $\left[\Gamma^{(A)}\right]^{\mu a}$, its mathematical expression, using the standard Feynman rules, is:
\begin{eqnarray}
    i\left[\Gamma^{(A)}\right]^{\mu a} &=& g^3 T^b T^a T^b \int \dfrac{d^Dl}{(2\pi)^D} \gamma^\alpha \dfrac{\slashed{p}^\prime+\slashed{l}+m}{(p^\prime+l)^2-m^2}  \gamma^\mu \dfrac{\slashed{p}+\slashed{l}+m}{(p+l)^2-m^2} \gamma^\beta \Delta_{\alpha \beta}(l) \,, \label{Vertex_l_Abel}
\end{eqnarray}
where the gluon propagator in an arbitrary covariant gauge reads as
\begin{eqnarray}
    \Delta_{\alpha \beta}(l) = \dfrac{1}{l^2} \left( g_{\alpha \beta} - \xi \dfrac{l_\alpha l_\beta}{l^2} \right) \,.
\end{eqnarray}
We can use the identities in Eqs.~\eqref{PropVertFO},~\eqref{Def_SO} 
and~\eqref{CAIdentity} to combine the first and second-order formalisms and rewrite the abelian part of the vertex function in Eq.~\eqref{Vertex_l_Abel} as
\begin{eqnarray}
    \left[\Gamma^{(A)}\right]^{\mu a} &=& g^3 \overline{C}\, T^a \int \dfrac{d^Dl}{i(2\pi)^D} \gamma^\alpha  \left( \dfrac{ B_{p+l,k}^\mu B_{p,l}^\beta}{D_{p^\prime+l} D_{p+l}} - \dfrac{B_{p+l,k}^\mu C_p^\beta}{D_{p^\prime+l} D_{p+l}} - \dfrac{\gamma^\mu \gamma^\beta}{D_{p^\prime+l}}  \right) \Delta_{\alpha \beta}(l) \, , \label{Vertex_l_Abel_combFS}
\end{eqnarray}
with $\overline{C} = C_F - C_A/2$, and where $C_F$ and $C_A$ are, respectively, the eigenvalues of the quadratic Casimir of $SU(N)$ in the fundamental and adjoint representations, respectively:
\begin{eqnarray}
    C_F = \dfrac{N^2-1}{2N} \,, \qquad C_A = N \,.
\end{eqnarray}
Now, the vector $B^\mu_{p+l,k}$ in Eq.~\eqref{Vertex_l_Abel_combFS} has clear separation into transverse and longitudinal components with respect to the external gluon momentum $k^\mu$. The longitudinal component corresponds to $(p+p^\prime+2l)^\mu$, while the transverse component corresponds to $\sigma^{\mu\nu}k_\nu$. This separation of longitudinal and transverse components remains after we commute $\gamma^\alpha$ with $B^{\mu}_{p+l,k}$, since:
\begin{eqnarray}
    \gamma^\alpha B_{p+l,k}^\mu &=& B_{p+l,k}^\mu \gamma^\alpha + 2(\gamma^\mu k^\alpha - g^{\mu\alpha} \slashed{k} ) \,, \nonumber \\ 
    &=& (p+p^\prime+2l)^\mu \gamma^\alpha - \sigma^{\mu\beta} k_\beta \gamma^\alpha + 2(\gamma^\mu k^\alpha - g^{\mu\alpha} \slashed{k} ) \,. \label{DefCommGandB}
\end{eqnarray}
Here, the transverse components correspond to the second and third terms in the second line of Eq.~\eqref{DefCommGandB}, while the longitudinal component corresponds to the first term. Thus, after commuting $\gamma^\alpha$ with $B^{\mu}_{p+l,k}$ in Eq.~\eqref{Vertex_l_Abel_combFS}, and contracting the Lorentz indices with the gluon propagator, we obtain the following natural decomposition of the abelian part of the quark-gluon vertex into its transverse and longitudinal components:
\begin{eqnarray}
    \left[\Gamma^{(A)}\right]^{\mu a} &=&  \dfrac{g^3 \overline{C}\, T^a}{(4\pi)^{\frac{D}{2}}}  \left( \left[\Gamma_L^{(A)}\right]^{\mu} + \left[\Gamma_T^{(A)}\right]^{\mu} \right) \,,    
\end{eqnarray}
where
\begin{eqnarray}
    \left[\Gamma_L^{(A)}\right]^{\mu} &=& \int \dfrac{d^Dl}{i \pi^{\frac{D}{2}} } \left\{ \dfrac{(p+p^\prime+2l)^\mu}{D_{p^\prime+l}D_{p+l} l^2} \left[  2 \slashed{p} + \left( 2 - D - \xi - 2 \xi \dfrac{ l \cdot p }{l^2} \right) \slashed{l} + (\xi-D)(\slashed{p}-m) \right] + \dfrac{(D-2-\xi)}{D_{p^\prime+l}l^2} \gamma^\mu  \right. \nonumber \\
    && \left. +  \dfrac{ 2 \xi l^\mu}{D_{p^\prime+l}l^4} \slashed{l} \right\} \,, \nonumber \\  \nonumber \\
    \left[\Gamma_T^{(A)}\right]^{\mu} &=& \int \dfrac{d^Dl}{i \pi^{\frac{D}{2}} } \dfrac{1}{D_{p^\prime+l}D_{p+l} l^2} \left\{ \sigma^{\mu\nu} k_\nu \left[   \left( D -2 + \xi + 2 \xi \dfrac{ l \cdot p }{l^2} \right) \slashed{l} -2 \slashed{p} + (D-4-\xi)(\slashed{p}-m) \right]  \right. \nonumber \\
    && \left. + \dfrac{2\xi}{l^2} ( k \cdot l \gamma^\mu - l^\mu \slashed{k}   ) \slashed{l} (\slashed{p}-m) + 4 \left[  k \cdot (p+l) \gamma^\mu - (p+l)^\mu \slashed{k} + ( k^\mu  - \gamma^\mu \slashed{k} ) \slashed{l} \right] \right. \nonumber \\
    && \left. - 2 \xi \left(1+2 \dfrac{l \cdot p}{l^2} \right) (k \cdot l \gamma^\mu - l^\mu \slashed{k})  \right\} \,. \label{LongTransAbelianVert}
\end{eqnarray}
To evaluate the tensor integrals of Eq.~\eqref{LongTransAbelianVert}, we can use the tensor reduction formulas given in Eqs.~\eqref{TI_1} and~\eqref{TI_2} to rewrite the one-loop tensor integrals in terms of the scalar integrals $J_2^D(a_1,a_2,a_3)$. Thus, from these formulas, the longitudinal $\left[\Gamma_L^{(A)}\right]^{\mu}$ and transverse $\left[\Gamma_T^{(A)}\right]^{\mu}$ components of the abelian part of the quark-gluon vertex in Eqs.~\eqref{LongTransAbelianVert} acquire the form 
\begin{eqnarray}
    \left[\Gamma_L^{(A)}\right]^{\mu} &=& (p+p^\prime)^\mu \left\{ \left[ (2-D)J_2^{D+2}(2,1,1) + \xi(p^2-m^2) J_2^{D+2}(2,1,2) - \xi J_2^{D+2}(2,0,2)  \right] \slashed{p} \right. \nonumber \\
    && \left. + \left[ (2-D)J_2^{D+2}(1,2,1) + \xi(p^2-m^2) J_2^{D+2}(1,2,2) \right] \slashed{p}^\prime + \left[ m(D-\xi) + (2-D+\xi)\slashed{p} \right] J_2^D(1,1,1) \right\} \nonumber \\
    && + 2 \left[ m(D-\xi) + (2-D+\xi)\slashed{p} \right] \left[ p^\mu J_2^{D+2}(1,2,1) + p^{\prime \mu} J_2^{D+2}(2,1,1) \right] + (D-2)\left[ \gamma^\mu J_2^{D+2}(1,1,1) \right. \nonumber \\
    && \left. - 2 \left( (2p^\mu J_2^{D+4}(1,3,1) + p^{\prime \mu} J_2^{D+4}(2,2,1) ) \slashed{p} + ( 2 p^{\prime \mu} J_2^{D+4}(3,1,1) + p^\mu J_2^{D+4}(2,2,1) ) \slashed{p}^\prime \right) \right] \nonumber \\
    && + (D-\xi-2) J_2^{D}(1,0,1) \gamma^\mu + \xi (m^2-p^2) \left[ J_2^{D+2}(1,1,2) \gamma^\mu - 2 \left( ( 2p^\mu J_2^{D+4}(1,3,2) + p^{\prime \mu} J_2^{D+4}(2,2,2) ) \slashed{p} \right. \right. \nonumber \\ 
    && \left. \left. + ( 2 p^{\prime \mu} J_2^{D+4}(3,1,2) + p^\mu J_2^{D+4}(2,2,2) ) \slashed{p}^\prime \right) \right] \,, \label{abelianl}    \\
    \left[\Gamma_T^{(A)}\right]^{\mu} &=& \sigma^{\mu\nu} k_\nu \left\{ \left[ (4-D+\xi)m + (D-6-\xi)\slashed{p} \right] J_2^{D}(1,1,1) + (D-2)\left[ \slashed{p} J_2^{D+2}(1,2,1) + \slashed{p}^\prime J_2^{D+2}(2,1,1)  \right] \right. \nonumber \\
    && \left. + \xi(m^2-p^2)\left[ \slashed{p} J_2^{D+2}(1,2,2) + \slashed{p}^\prime J_2^{D+2}(2,1,2) \right] + \xi \slashed{p}^\prime J_2^{D+2}(2,0,2) \right\} + 4 J_2^{D}(1,1,1) ( k \cdot p \gamma^\mu - p^\mu \slashed{k}  ) \nonumber \\
    &&+ 2 \xi \big\{ \gamma^\mu \slashed{p} \left[ 2 k \cdot p J_2^{D+4}(1,3,2) + k \cdot p^\prime J_2^{D+4}(2,2,2) \right] - \slashed{k} \slashed{p} \left[ 2p^\mu J_2^{D+4}(1,3,2) + p^{\prime\mu} J_2^{D+4}(2,2,2) \right]  \nonumber \\
    && + \gamma^\mu \slashed{p}^\prime \left[ 2 k \cdot p^\prime J_2^{D+4}(3,1,2) + k \cdot p J_2^{D+4}(2,2,2) \right] - \slashed{k} \slashed{p}^\prime \left[ 2p^\mu J_2^{D+4}(1,3,2) + p^{\prime\mu} J_2^{D+4}(2,2,2) \right] \nonumber \\
    && J_2^{D+2}(1,1,2)( k^\mu - \gamma^\mu \slashed{k} ) \big\} (\slashed{p}-m) + 4 \big[ \gamma^\mu \left( k \cdot p J_2^{D+2}(1,2,1) + k \cdot p^\prime J_2^{D+2}(2,1,1)  \right) \nonumber \\
    && - \slashed{k} \left(  p^\mu J_2^{D+2}(1,2,1) + p^{\prime\mu} J_2^{D+2}(2,1,1)  \right) \big] + 4 (k^\mu-\gamma^\mu \slashed{k})\left[ \slashed{p} J_2^{D+2}(1,2,1) + \slashed{p}^\prime J_2^{D+2}(2,1,1) \right] \nonumber \\
    && - 2 \xi J_2^{D+2}(2,0,2) \left( k \cdot p^\prime \gamma^\mu - p^{\prime \mu} \slashed{k}  \right) - 2 \xi (m^2-p^2)\big[ \gamma^\mu \left( k \cdot p J_2^{D+2}(1,2,2) + k \cdot p^\prime J_2^{D+2}(2,1,2) \right) \nonumber \\
    && - \slashed{k} \left( p^\mu J_2^{D+2}(1,2,2) + p^{\prime \mu} J_2^{D+2}(2,1,2) \right)  \big] \,. \label{abeliant}
\end{eqnarray}
These expressions are equivalent to the results presented in Ref.~\cite{QEDVer}, with the corresponding change of conventions. 

\subsection{Off-shell non-abelian graph}

The non-abelian contribution $\left[\Gamma^{(NA)}\right]^{\mu a}$ to the quark-gluon vertex within the first-order formalism, see Fig.~\ref{NonAbelian_graph}, reads as,
\begin{eqnarray}
\left[\Gamma^{(NA)}\right]^{\mu a} = g^3 f^{abc} T^c T^b  \int \dfrac{d^Dl}{(2\pi)^D} \gamma^{\nu^\prime} \left( \dfrac{-\slashed{l}+m}{l^2-m^2} \right) \gamma^{\rho^\prime} \Delta_{\rho \rho^\prime}(p+l) \Delta_{\nu\nu^\prime}(p^\prime+l) V_{\rho \nu \mu}(p+l,-p^\prime-l,k) \,, \label{NA_vertex_1}
\end{eqnarray}
where the three-gluon vertex $V_{\alpha\beta\mu}$ is given by
\begin{eqnarray}
    V_{\alpha\beta\mu}(r,s,t) &=& g^{\alpha\mu}(t-r)^{\beta} + g^{\alpha\beta} (r-s)^\mu + g^{\beta\mu}(s-t)^\alpha \,. \label{gluon_3_vertex}
\end{eqnarray}
After performing the color algebra, and contracting the Lorentz indices of the quark and gluon propagators and the three-gluon vertex, the expression in Eq.~\eqref{NA_vertex_1} becomes
\begin{eqnarray}
    \left[\Gamma^{(NA)}\right]^{\mu a} &=& \dfrac{1}{2} g^3 C_A T^a \int \dfrac{d^Dl}{i(2\pi)^D} \gamma^{\nu^\prime} \left( \dfrac{-\slashed{l}+m}{q^{\prime 2} q^2 D_l} \right) \gamma^{\rho^\prime} \left\{ g^{\mu\rho^\prime} (k-q)^{\nu^\prime} - g^{\mu\nu^\prime}(k+q^\prime)^{\rho^\prime} + g^{\nu^\prime \rho^\prime}(q+q^\prime)^\mu \right. \nonumber \\
    && + \dfrac{\xi q^{\rho^\prime}}{q^2} \left[ g^{\mu\nu^\prime} q \cdot (k+q^\prime) - k^{\nu^\prime} q^\mu - q^{\nu^\prime} q^{\prime \mu}  \right] + \dfrac{\xi q^{\prime \nu^\prime}}{q^{\prime 2}}\left[ g^{\mu\rho^\prime} q^\prime \cdot (q-k) + k^{\rho^\prime}q^{\prime \mu} - q^\mu q^{\prime \rho^\prime} \right] \nonumber \\
    && \left. + \dfrac{\xi^2 q^{\rho^\prime}q^{\prime \nu^\prime}}{q^{\prime 2} q^2} \left( k \cdot q^\prime q^\mu - k \cdot q q^{\prime \mu} \right) \right\} \,, \label{NA_vertex_2}
\end{eqnarray}
where $q=p+l$ and $q^\prime=p^\prime+l$.
Similarly to the abelian contribution, we combine the first and second-order formalism by rewriting
\begin{eqnarray}
    \gamma^{\nu^\prime} (-\slashed{l}+m) \gamma^{\rho^\prime} &=& \gamma^{\nu^\prime} A_{-l,0}^{\rho^\prime} = \gamma^{\nu^\prime} B_{-l,0}^{\rho^\prime} - \gamma^{\nu^\prime} C^{\rho^\prime}_{-l} \nonumber \\
    && \hspace{-40pt} =  -2 \gamma^{\nu^\prime} l^{\rho^\prime} + ( \sigma^{\nu^\prime\rho^\prime} + g^{\nu^\prime\rho^\prime} )(\slashed{l}+m). \label{nonabelian_dec} 
\end{eqnarray}
Using this identity, Eq.~\eqref{NA_vertex_2} can be cast in the following form:
\begin{eqnarray}
\left[\Gamma^{(NA)}\right]^{\mu a} &=& \dfrac{1}{2(4\pi)^{\frac{D}{2}}} g^3 C_A T^a \left( \left[\Gamma_L^{(NA)}\right]^{\mu} + \left[\Gamma_T^{(NA)}\right]^{\mu} \right) \,,
\end{eqnarray}
where the longitudinal $\left[\Gamma_L^{(NA)}\right]^{\mu}$ and transverse $\left[\Gamma_T^{(NA)}\right]^{\mu}$ components of the non-abelian graph that arise from the decomposition \eqref{nonabelian_dec} read as follows:
\begin{eqnarray}
    \left[\Gamma_L^{(NA)}\right]^{\mu} &=& \int \dfrac{d^Dl}{i\pi^{\frac{D}{2}}} \dfrac{1}{q^{\prime 2}q^2 D_l } \Big\{ (q+q^\prime)^\mu \left[ (D-3-\xi) \slashed{l} + m (D-1-\xi) \right] - 2(\slashed{k}-\slashed{q})l^\mu + 2 \gamma^\mu l \cdot (k+q^\prime) \nonumber \\
    && + \dfrac{\xi}{q^{\prime 2}} \slashed{q}^\prime \left[ 2 \left( q^\prime \cdot (k-q) l^\mu - l \cdot k q^{\prime \mu} + l \cdot q^\prime q^\mu \right) + \left( (q^2-k^2)\gamma^\mu + q^{\prime \mu} \slashed{k} \right)(\slashed{l}+m) \right] \nonumber \\
    && - \dfrac{\xi}{q^2} \left[ 2 l \cdot q \left( q \cdot (k+q^\prime)\gamma^\mu - \slashed{k} q^\mu - \slashed{q} q^{\prime \mu} \right) + \left( (k^2-q^{\prime 2}) \gamma^\mu + q^\mu \slashed{k} \right) \slashed{q} (\slashed{l}+m) \right] \Big\} \,,  \nonumber \\ \nonumber \\
    \left[\Gamma_T^{(NA)}\right]^{\mu} &=& -\int \dfrac{d^Dl}{i\pi^{\frac{D}{2}}} \dfrac{1}{q^{\prime 2}q^2 D_l} \Big\{  3 \sigma^{\mu\nu} k_\nu (\slashed{l}+m) + \dfrac{\xi^2}{q^{\prime 2} q^2} \slashed{q}^\prime \left[ 2 l \cdot q - \slashed{q} (\slashed{l}+m) \right]\left( q^\mu k \cdot q^\prime - q^{\prime \mu} k \cdot q \right)  \Big\} \,. \label{NA_LT_1}
\end{eqnarray}
\end{widetext}
Since the external gluon is not attached to a fermion line in the non-abelian case, the combination of the second-order formalism with the first-order formalism does not have a cleaner separation of longitudinal and transverse components as in the abelian diagram. However, it does provide a compact tensor structure for applying the tensor reduction formulas (Eqs.~\eqref{TI_1} and~\eqref{TI_2}). On using these formulas, the longitudinal and transverse components given in Eq.~\eqref{NA_LT_1} become
\begin{eqnarray}
    \left[\Gamma_L^{(NA)}\right]^{\mu} &=& \sum_{i=1}^{12} \alpha_i h^\mu_i \,,  \nonumber \\ \nonumber \\
    \left[\Gamma_T^{(NA)}\right]^{\mu} &=& \sum_{i=1}^8 \beta_i T^\mu_i \,, \label{Vertex_NA}
\end{eqnarray}
where
\begin{align}
    h_1^\mu &= \gamma^\mu, &\quad 
    h_2^\mu &= p^\mu, &\quad 
    h_3&=p^{\prime\mu}, \nonumber \\
    h_4^\mu&=\gamma^\mu \slashed{p}, &\quad
    h_5^\mu&=\gamma^\mu \slashed{p}^\prime, &\quad 
    h_6^\mu&=p^\mu \slashed{p}, \nonumber \\
    h_7^\mu&=p^\mu \slashed{p}^\prime, &\quad 
    h_8^\mu&=p^{\prime\mu} \slashed{p}, &\quad
    h_9^\mu&=p^{\prime \mu} \slashed{p}^\prime, \nonumber \\
    h_{10}^\mu&=p^\mu \slashed{p}^\prime \slashed{p}, &\quad h_{11}^\mu&=p^{\prime \mu} \slashed{p}^\prime \slashed{p}, &\quad h_{12}^\mu&=\gamma^\mu \slashed{p}^\prime \slashed{p}, \label{h_basis} \nonumber \\
\end{align}
and
\begin{align}
    T_1^\mu&= p^\prime \cdot k p^\mu - p \cdot k p^{\prime \mu} \,, \nonumber  \\ 
    T_2^\mu&= (p^\prime \cdot k p^\mu - p \cdot k p^{\prime \mu}) (\slashed{p}+\slashed{p}^\prime) \,, \nonumber \\
    T_3^\mu&= k^2 \gamma^\mu - k^\mu \slashed{k} \,, \nonumber \\
    T_4^\mu&= -(p^\prime \cdot k p^\mu - p \cdot k p^{\prime \mu})\sigma_{\nu\lambda}p^{\prime \nu} p^\lambda \,, \nonumber \\
    T_5^\mu&=\sigma^{\mu\nu} k_\nu \,, \nonumber \\
    T_6^\mu&=\gamma^\mu(p^{\prime 2}-p^2)-(p+p^\prime)^\mu \slashed{k} \,, \nonumber \\
    T_7^\mu&= \frac{1}{2}(p^2-p^{\prime 2})[(p+p^\prime)^\mu-\gamma^\mu(\slashed{p}+\slashed{p}^\prime)] \nonumber \\
    &-(p+p^\prime)^\mu \sigma^{\nu\lambda} p^{\prime \nu} p^\lambda \,, \nonumber \\
    T_8^\mu&= \gamma^\mu \sigma_{\nu\lambda}p^{\prime\nu}p^\lambda - p^{\prime\mu}\slashed{p} + p^\mu \slashed{p}^\prime \,. \label{TBallChiu}
\end{align}
The coefficients $\alpha_i$ and $\beta_i$ which form part of Eq.~\eqref{Vertex_NA} are explicitly collected in Appendix~\ref{coefficients} in terms of scalar integrals with shifted dimensions and higher powers in the propagators. 

The quark-gluon vertex component $\left[\Gamma_T^{(NA)}\right]^{\mu}$ in Eq.~\eqref{Vertex_NA} is written in terms of the conventional Ball-Chiu transverse basis $T^\mu_i$, used extensively in the literature for some decades (see, for example, Refs.~\cite{Ball_chiu,Davydychev,Pennington}), so that we can directly compare with the results presented in previous works. Notice that $\left[\Gamma_L^{(NA)}\right]^{\mu}$ contains all possible tensor structures for the quark-gluon vertex, which implies that there are more hidden transverse components in this term than in its abelian counterpart $\left[\Gamma_T^{(A)}\right]^{\mu}$ given in  Eq.~\eqref{abelianl}. 

\subsection{Ball Chiu decomposition}

A commonly adopted approach for expressing the complete quark-gluon vertex is by decomposing it in the Ball-Chiu longitudinal and transverse components, Refs.~\cite{Ball_chiu,Davydychev,Pennington},
\begin{eqnarray}
    \Gamma^\mu = \sum_{i=1}^4 \lambda_i L_i^\mu + \sum_{i=1}^{8} \tau_i T_i^\mu \,,
\end{eqnarray}
where the transversal basis $T_i^\mu$ is given in Eq.~\eqref{TBallChiu}, and the longitudinal basis $L_i^\mu$ corresponds to the following choice of vectors
\begin{eqnarray}
    L_1^\mu &=& \gamma^\mu \,, \nonumber \\
    L_2^\mu &=& (p+p^\prime)^\mu (\slashed{p}+\slashed{p}^\prime) \,, \nonumber \\
    L_3^\mu &=& - (p+p^\prime)^\mu \,, \nonumber \\
    L_4^\mu &=& -\sigma^{\mu\nu}(p+p^\prime)_\nu \,.
\end{eqnarray}
In Appendix~C of Ref.~\cite{QEDVer}, we obtain the coefficients $\lambda_i$ and $\tau_i$ for the one-loop abelian contribution to the quark-gluon vertex in terms of scalar integrals with shifted dimensions and higher power in the propagators from the  results equivalent to the expressions given in Eqs.~\eqref{abelianl} and~\eqref{abeliant}. For details of this calculation, we refer the reader to that work. For the non-abelian contribution, the coefficients $\lambda_i^{(NA)}$ and $\tau_i^{(NA)}$, where the superscript label stands for non-abelian, read as
\begin{eqnarray}
    \lambda_1 &=& \alpha_1 - \frac{1}{2}(\alpha_6+\alpha_7) k \cdot p  + \frac{1}{2} (\alpha_8+\alpha_9) k \cdot p^\prime \nonumber \\
    && + \frac{1}{2}(k^2-2p^2) \alpha_{12} \,, \nonumber \\
    \lambda_2 &=& \frac{1}{2(p^2-p^{\prime 2})}\big[ -(\alpha_6-\alpha_7) k \cdot p + (\alpha_8-\alpha_9) k \cdot p^\prime \nonumber \\
    && + \alpha_{12} k^2  \big] \,, \nonumber \\
    \lambda_3 &=& \frac{1}{p^2-p^{\prime 2}}\big[  k \cdot p (\alpha_2+\alpha_4-p \cdot p^\prime \alpha_{10}) \nonumber \\
    && - k \cdot p^\prime (\alpha_3+\alpha_5-p \cdot p^\prime \alpha_{11}) \big] \,, \nonumber \\
    \lambda_4 &=& \frac{1}{2}\big[ -\alpha_4 + \alpha_5 + k \cdot p \, \alpha_{10} - k \cdot p^\prime\, \alpha_{11} \big] \,, \label{lamdaNA}
\end{eqnarray}
\begin{eqnarray}
    \tau_1 &=& - \frac{1}{p^2-p^{\prime 2}}\big[ \alpha_2 + \alpha_3 + \alpha_4 + \alpha_5 \nonumber \\
    && - p \cdot p^\prime (\alpha_{10}+\alpha_{11}) \big] + \beta_1 \,, \nonumber \\
    \tau_2 &=& - \frac{1}{2(p^2-p^{\prime 2})} \big[ \alpha_6 -\alpha_7 +\alpha_8 - \alpha_9 + 2 \alpha_{12} \big] \nonumber \\
    && + \beta_2 \,, \nonumber \\
    \tau_3 &=& -\frac{1}{4} \big[ \alpha_6 + \alpha_7 + \alpha_8 + \alpha_9 \big] + \beta_3 \,, \nonumber \\
    \tau_4 &=& - \frac{1}{p^2-p^{\prime 2}} \big[ \alpha_{10} + \alpha_{11} \big] + \beta_4 \,, \nonumber \\
    \tau_5 &=& - \frac{1}{2} \big[ \alpha_4 + \alpha_5 \big] + \beta_5 \,, \nonumber \\
    \tau_6 &=& \frac{1}{4} \big[ \alpha_6 + \alpha_7 - \alpha_8 - \alpha_9 - 2 \alpha_{12} \big] + \beta_6 \,, \nonumber \\
    \tau_7 &=& - \frac{1}{p^2-p^{\prime 2}} \big[ k \cdot p\, \alpha_{10} - k \cdot p^\prime \alpha_{11} \big] + \beta_7 \,, \nonumber \\
    \tau_8 &=& - \alpha_{12} + \beta_8 \,, \label{tauNA}
\end{eqnarray}
where the coefficients $\alpha_i$ and $\beta_i$ are given in Appendix~\ref{coefficients}. 

After we apply the dimensional reduction formula of Eq.~\eqref{DimRedJi}, and employ the integration by parts technique to rewrite the scalar integrals of Appendix~\ref{coefficients} into the master integrals of Eq.~\eqref{MasterInt}, we obtain precisely the same results as presented in Ref.~\cite{Davydychev} for the non-abelian coefficients $\lambda_i^{(NA)}$ and $\tau_i^{(NA)}$ of Eqs.~\eqref{lamdaNA} and~\eqref{tauNA}. Notice that $p_1$, $p_2$ and $p_3$ of Ref.~\cite{Davydychev} correspond to $p^\prime$, $p$ and $k$, respectively, in this work.   

\section{On-shell quark-gluon vertex: A scalar-spin interaction analysis} \label{Onshell_section}

For the case of on-shell quarks, we can perform a complete transition from first-order amplitudes to their 
second-order version. Once we have a second-order version of the amplitude, we can analyze and classify its spin and scalar interactions. To illustrate this kind of analysis, we now compute the contributions to the quark magnetic and chromo-magnetic dipole moments that arise from different scalar and spin interactions. In the following, we fix $\xi=0$ in order to simplify the calculations. 

\subsection{On-shell abelian graph}

According to the first-order rules, the product of the on-shell spinors and the one-loop $\left[\Gamma^{(A)}\right]^{\mu a}$ vertex reads as,
\begin{eqnarray}
    && \bspinor \left[\Gamma^{(A)}\right]^{\mu a} \spinor = \bspinor \left[ g^3 \overline{C} T^a  \int \dfrac{d^Dl}{i(2\pi)^D} \gamma^\alpha \right. \nonumber \\
    && \hspace*{10pt} \left. \times \dfrac{\slashed{p}^\prime+\slashed{l}+m}{(p^\prime+l)^2-m^2} \gamma^\mu \dfrac{\slashed{p}+\slashed{l}+m}{l^2[(p+l)^2-m^2]} \gamma_\alpha \right] \spinor, \label{qed_vert_spinors}
\end{eqnarray}
where the spinors $\bspinor$ and $\spinor$ satisfy the on-shell conditions,
\begin{align}
    (\slashed{p}-m)u_s(p) = 0 \,, \quad \bspinor(\slashed{p}^\prime-m) = 0 \,. \label{onshell_spinors}
\end{align}
Now, for a complete description of the expression in Eq.~\eqref{qed_vert_spinors} within the second-order formalism, we introduce a prefactor, as discussed in Ref.~\cite{Morgan}, before the first $\gamma^\alpha$ matrix using the following identity
\begin{eqnarray}
    \bspinor = \bspinor \dfrac{(\slashed{p}^\prime + m)}{2m} \,. \label{iden_additional_prop}
\end{eqnarray}
Thus, the expression in \eqref{qed_vert_spinors} becomes
\begin{eqnarray}
    && \hspace{-5mm} \bspshort \qedvert \spshort = \overline{C}\, \bspshort \left[ \dfrac{g^3}{2m} \int \dfrac{d^Dl}{i(2\pi)^D} (\slashed{p}^\prime + m) \right. \nonumber \\
    && \hspace{-5mm} \hspace*{10pt} \left. \times \gamma^\alpha \dfrac{\slashed{p}^\prime+\slashed{l}+m}{(p^\prime+l)^2-m^2} \gamma^\mu \dfrac{\slashed{p}+\slashed{l}+m}{l^2[(p+l)^2-m^2]} \gamma_\alpha \right] \spshort \,,
\end{eqnarray}
where we have suppressed the color factor $T^a$, and the momentum dependence of the spinors $\spinor$ and $\bspinor$ for notational simplicity. 
From Eq.~\eqref{PropVertFO}, the expression above can be rewritten as
\begin{eqnarray}
     \bspshort \qedvert \spshort &=& \overline{C}\, \bspshort \left[ \dfrac{g^3}{2m} \int \dfrac{d^Dl}{i(2\pi)^D} \dfrac{ 1 }{l^2 D_{p^\prime+l} D_{p+l} } \right. \nonumber \\
     && \times A^\alpha_{p^\prime+l,-l} A^\mu_{p^\prime,l} A_{\alpha\,p,l} \left. \right] \spshort \,, \label{qed_vert_spinors2}
\end{eqnarray}
After employing the definitions given in Eq.~\eqref{Def_SO}, the identity in Eq.~\eqref{CAIdentity}, and the spinor on-shell conditions given in Eq.~\eqref{onshell_spinors}, the expression in Eq.~\eqref{qed_vert_spinors2} becomes
\begin{eqnarray}
     && \bspshort \qedvert \spshort = \overline{C}\, \bspshort \left\{ \dfrac{g^3}{2m} \int \dfrac{d^Dl}{i(2\pi)^D} \dfrac{1 }{l^2 D_{p^\prime+l} D_{p+l} } \right. \nonumber \\
    && \hspace*{10pt} \times \left[  B^\alpha_{p^\prime+l,-l} B^\mu_{p+l,k} B_{\alpha\,p,l} - D_{p+l} B_{\alpha\,p^\prime+l,-l} ( \sigma^{\mu\alpha} + g^{\mu\alpha} ) \right. \nonumber \\
    && \hspace*{10pt} \left. - D_{p^\prime+l} (\sigma^{\alpha \mu}+g^{\mu\alpha}) B_{\alpha\,p,l}
    \right] \bigg\} \spshort \,, \label{qed_vert_SO}
\end{eqnarray}
which now contains only second-order scalar and spin (vertices) interactions that can be grouped as
\begin{eqnarray}
     && \bspshort \qedvert \spshort = \dfrac{g\overline{C}}{2m} \bspshort \left(  \left[\Gamma^A_{\text{sc-sc-sc}}\right]^{\mu} \right. \nonumber \\
     && \left. \hspace{20pt} + \left[\Gamma^A_{\text{sp-sc-sc}}\right]^{\mu} + \left[\Gamma^A_{\text{sp-sp-sc}}\right]^{\mu} + \left[\Gamma^A_{\text{sc-sc-sp}}\right]^{\mu}  \right. \nonumber \\
     && \left. \hspace{20pt} + \left[\Gamma^A_{\text{sp-sc-sp}}\right]^{\mu}  + \left[\Gamma^A_{\text{sp-sp-sp}}\right]^{\mu} + \left[\Gamma^A_{\text{sc-sc}}\right]^{\mu}   \right. \nonumber \\
     && \left. \hspace{20pt} + \left[\Gamma^A_{\text{sp-sc}}\right]^{\mu} + \left[\Gamma^A_{\text{sc-sp}}\right]^{\mu} + \left[\Gamma^A_{\text{sp-sp}}\right]^{\mu}  \right) \spshort \,. \label{qed_vert_SO_2}
\end{eqnarray}
Here, the second-order vertex functions $\left[\Gamma^A_{a}\right]^\mu$ are listed in Appendix~\ref{SO_vert_fun}. The subscript label $a$ in the vertex functions denotes the type of interactions between the gluons and the fermion line. For example, $\left[\Gamma^A_{\text{sc-sc-sp}}\right]^\mu$ describes a second-order diagram where the virtual gluon with momentum $l$ performs two scalar interactions with the fermion line, while the external gluon with momentum $k$ performs a spin interaction (see Fig.~\ref{Gamma_scscsp}). Similarly, $\left[\Gamma^A_{\text{sc-sp}}\right]^\mu$ denotes a scalar interaction of the virtual gluon, and a seagull spin interaction of the external gluon (see Fig.~\ref{Gamma_scsp}).

We can formally get the QED case from the expressions above by setting $g=e$, $C_A=0$ and $C_F=1$. Thus, we can obtain the one-loop second-order contributions to the $g$-factor of the electron from Eq.~\eqref{qed_vert_SO_2}.  
Most generally, the on-shell QED vertex can be written as
\begin{eqnarray}
    && \bspinor \qedvert \spinor \nonumber \\
    && \hspace{10pt} = e \bspshort \left[ \dfrac{F^A_1(k^2)}{2m} (p+p^\prime)^\mu - \dfrac{F^A_{\text{sp}}(k^2)}{2m} \sigma^{\mu\nu} k_\nu \right] \,, \label{QED_vertex}
\end{eqnarray}
with $F^A_{\text{sp}}(k^2)$ given by
\begin{eqnarray}
    F^A_{\text{sp}}(k^2) = F^A_1(k^2) + F^A_2(k^2) \,.
\end{eqnarray}
Here, $F^A_1(k^2)$ and $F^A_2(k^2)$ are the Dirac and Pauli form factors of the electron, respectively.
Using the results of the Appendix~\ref{SO_vert_fun}, the one-loop form factors in the Feynman gauge read as
\begin{eqnarray}
    F^A_1(k^2) &=& \frac{e^2}{(4\pi)^{\frac{D}{2}}}\Big\{ 
    (D-2) \big[ J_2^{D+2}(1,1,1) + J_2^D(1,1,0) \nonumber \\
    && \hspace{-30pt} -4 m^2 (J_2^{D+4}(2,2,1)+2 J_2^{D+4}(3,1,1)) \big] \nonumber \\
    && \hspace{-30pt} +4 p \cdot p^\prime \big[J_2^D(1,1,1)-2 J_2^{D+2}(2,1,1) \big] \Big\} \,, \nonumber \\
    F^A_{\text{sp}}(k^2) &=& \frac{e^2}{(4\pi)^{\frac{D}{2}}}\Big\{
    4 p \cdot p^\prime J_2^D(1,1,1) \nonumber \\
    && \hspace{-30pt} -8 \left(m^2+ p \cdot p^\prime \right) J_2^{D+2}(2,1,1) +(D-2) \big[ J_2^D(1,1,0) \nonumber \\
    && \hspace{-30pt} +J_2^{D+2}(1,1,1) \big] \Big\} \,, \nonumber \\
    F^A_2(k^2) &=& \frac{4m^2e^2}{(4\pi)^{\frac{D}{2}}}\Big\{ (D-2) \big[ J_2^{D+4}(2,2,1) \nonumber \\
    && \hspace{-30pt} +2J_2^{D+4}(3,1,1) \big] -2 J_2^{D+2}(2,1,1) \Big\} \,, \label{Abelian_FF}
\end{eqnarray}
which agree with our previous results reported in \cite{QEDVer,RQED}, and with the results of Ref.~\cite{Davydychev}.

From Eq.~\eqref{QED_vertex} one can deduce that the $g$-factor, which measures the strength of the interaction of the electron with a magnetic field, corresponds to 
\begin{eqnarray}
    g = 2 F^A_{\text{sp}}(0) = 2 + 2 F^A_2(0) \,,
\end{eqnarray}
where the renormalization condition $F^A_1(0) = 1$ is used, see for example Ref.~\cite{Peskin}.
The contribution of the second-order diagrams to the $g$-factor at one-loop (see Appendix~\ref{second_order_diagrams}) is summarized in the Table~\ref{g_contributions}, where the form factor $F_1(0)$ reads as
\begin{eqnarray}
    F^A_1(0) &=& \dfrac{\alpha}{4\pi} \left[ \dfrac{1}{\epsilon} + \ln\left(\dfrac{\bar{\mu}^2}{m^2}\right) - 2 \ln\left(\dfrac{m^2}{m_\gamma^2}\right) + 4 \right].\label{F1_0}
\end{eqnarray}
\begin{table}[h!]
\caption{\label{g_contributions} \justifying{
Contributions to $F^A_{\text{sp}}(0)$ that arise from the second-order diagrams displayed in Fig.~\ref{SO_diagrams_QED_vertex}, grouped according to each type of second-order interactions. }
}
\begin{ruledtabular}
\begin{tabular}{l l }
\textrm{Second-order interactions} & $4\pi \bar{F}_{\text{sp}}^a(0)/\alpha$  \\
\colrule
Scalar-scalar-spin& $ \dfrac{4\pi}{\alpha} F^A_1(0) $\\
Spin-spin-scalar& 0\\
Scalar-scalar-scalar& $0$ \\
Spin-spin-spin& $-\dfrac{1}{\epsilon}-\ln\left(\dfrac{\bar{\mu}^2}{m^2}\right)$ \\
Spin-scalar-scalar& $-\dfrac{1}{\epsilon}-\ln\left(\dfrac{\bar{\mu}^2}{m^2}\right)-1$\\
Spin-scalar-spin& $0$\\
Scalar-scalar& $0$ \\
Scalar-spin& $\dfrac{3}{2\epsilon} + \dfrac{3}{2} \ln\left(\dfrac{\bar{\mu}^2}{m^2}\right) + \dfrac{7}{2} $\\
Spin-scalar& $-\dfrac{1}{2\epsilon}-\dfrac{1}{2}\ln\left(\dfrac{\bar{\mu}^2}{m^2}\right)-\dfrac{1}{2}$\\
Spin-spin& $\dfrac{1}{\epsilon}+\ln\left(\dfrac{\bar{\mu}^2}{m^2}\right)$\\
\end{tabular}
\end{ruledtabular}
\end{table}

From Table~\ref{g_contributions} we can read off the following results:
\begin{itemize}
    \item If we sum the expressions in the second column, multiply the result by the factor $\alpha/(4\pi)$, and subtract $F_1(0)$, we obtain the result of Schwinger:
    \begin{equation*}
        F_2(0)= \dfrac{g-2}{2} =\dfrac{\alpha}{2\pi} + O(\alpha^2) \,.
    \end{equation*} 
    \item The pure scalar interactions, diagrams \ref{Gamma_scscsc} and \ref{Gamma_scsc}, as expected, do not contribute to the $g$-factor since they do not affect the spin of the electron.
    \item The diagrams \ref{Gamma_spspsc} and \ref{Gamma_spscsp}, as in the case of the pure scalar interaction diagrams, do not contribute either to the $g$-factor.
    \item The contribution to the $g$-factor that arise from the diagram \ref{Gamma_scscsp} is equal to the Dirac form factor $F^A_1(0)$.
    \item The diagrams \ref{Gamma_spspsp} and \ref{Gamma_spsp} with pure spin interactions do not play a relevant physical role in the contribution to the $g$-factor since they contain only a divergent term and a logarithmic term on the scale parameter $\mu$ which cancel with contributions arising from other diagrams.  
    \item From all second-order diagrams, the only diagrams that contain physical information about the anomalous magnetic moment at one-loop are \ref{Gamma_spscsc}, \ref{Gamma_scsp} and \ref{Gamma_spsc}. They have finite and $\mu$-scale independent contributions to the $g$-factor equal to $-\frac{\alpha}{4\pi}$, $\frac{7\alpha}{8\pi}$ and $-\frac{\alpha}{8 \pi}$ respectively. When these contributions are summed, Schwinger's result is recovered.  
\end{itemize}

\begin{widetext}

\begin{figure}[h!]
\begin{subfigure}[b]{0.45\textwidth}
\includegraphics[scale=0.35]{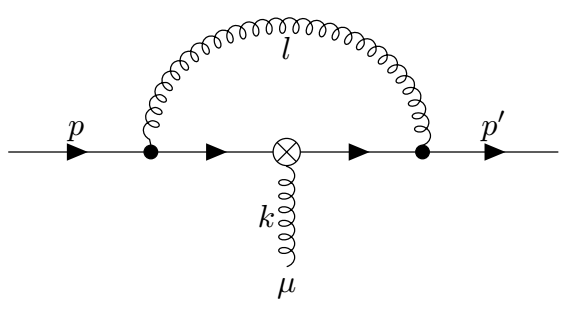}
\caption{Scalar-scalar-spin abelian interaction (sc-sc-sp).} \label{Gamma_scscsp}
\end{subfigure}
\begin{subfigure}[b]{0.45\textwidth}
\includegraphics[scale=0.35]{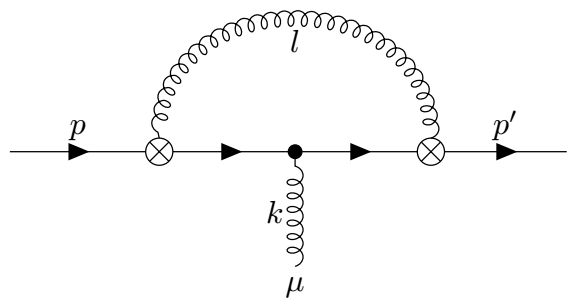}
\caption{Spin-spin-scalar abelian interaction (sp-sp-sc).} \label{Gamma_spspsc}
\end{subfigure} 
\begin{subfigure}[b]{0.45\textwidth}
\includegraphics[scale=0.35]{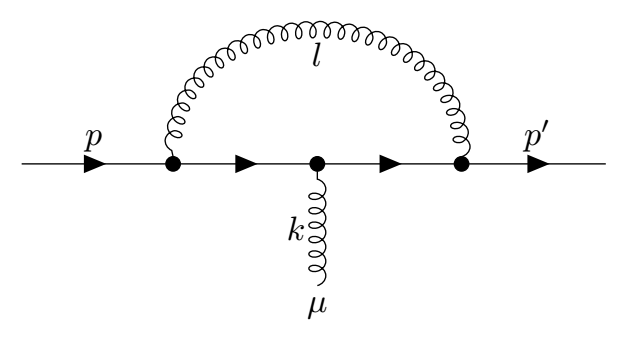}
\caption{Scalar-scalar-scalar abelian interaction (sc-sc-sc).} \label{Gamma_scscsc}
\end{subfigure}
\begin{subfigure}[b]{0.45\textwidth}
\includegraphics[scale=0.35]{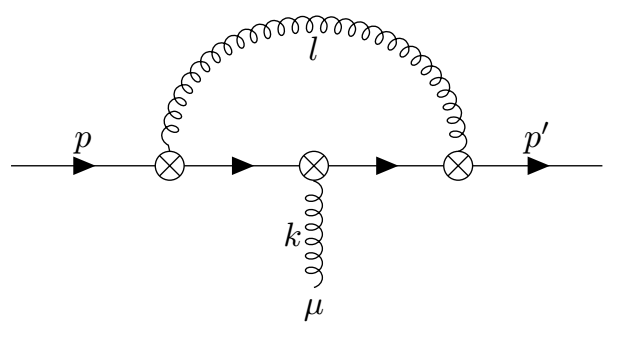}
\caption{Spin-spin-spin abelian interaction (sp-sp-sp).} \label{Gamma_spspsp}
\end{subfigure}
\begin{subfigure}[b]{0.45\textwidth}
\hspace*{-0.8cm} \includegraphics[scale=0.35]{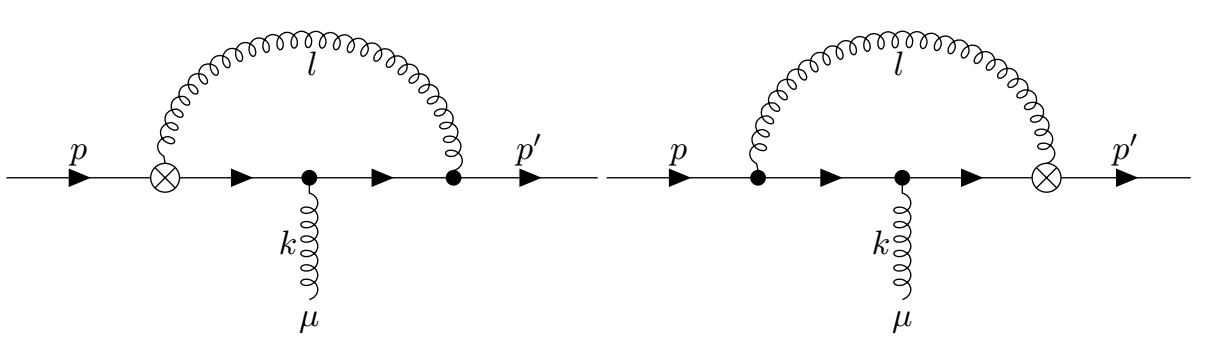}
\caption{Spin-scalar-scalar abelian interaction (sp-sc-sc).} \label{Gamma_spscsc}
\end{subfigure}
\begin{subfigure}[b]{0.45\textwidth}
\hspace*{0.1cm} \includegraphics[scale=0.35]{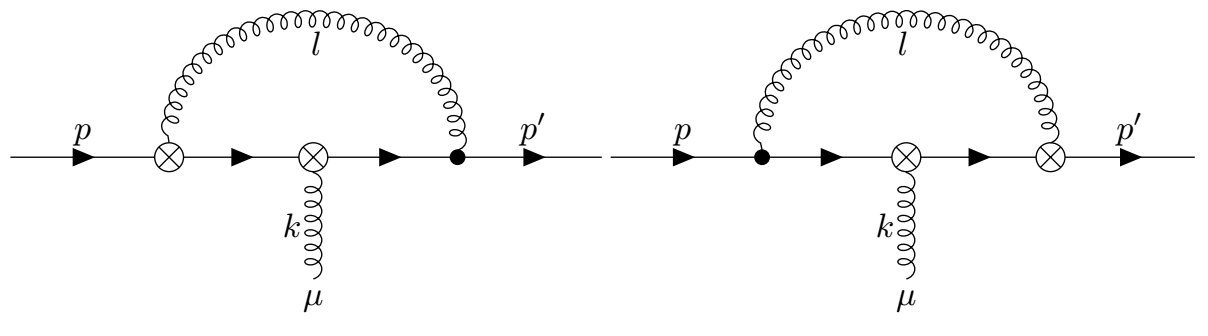}
\caption{Spin-scalar-spin abelian interaction (sp-sc-sp).} \label{Gamma_spscsp}
\end{subfigure}
    \begin{subfigure}[b]{0.45\textwidth}
    \includegraphics[scale=0.35]{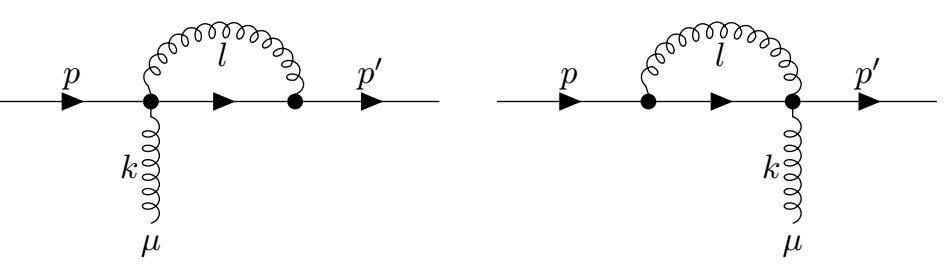}
    \caption{Scalar-scalar abelian interaction (sc-sc).} \label{Gamma_scsc}
\end{subfigure}
\begin{subfigure}[b]{0.45\textwidth}
    \includegraphics[scale=0.35]{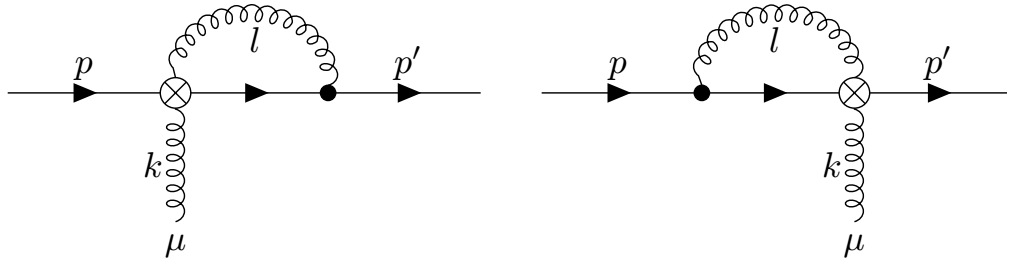}
    \caption{Scalar-spin abelian interaction (sc-sp).} \label{Gamma_scsp}
\end{subfigure}
\begin{subfigure}[b]{0.45\textwidth}
    \includegraphics[scale=0.35]{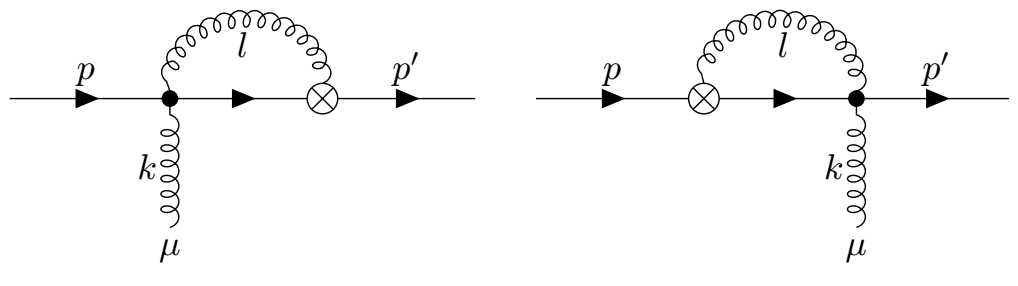}
    \caption{Spin-scalar abelian interaction (sp-sc).} \label{Gamma_spsc}
\end{subfigure}
\begin{subfigure}[b]{0.45\textwidth}
    \includegraphics[scale=0.35]{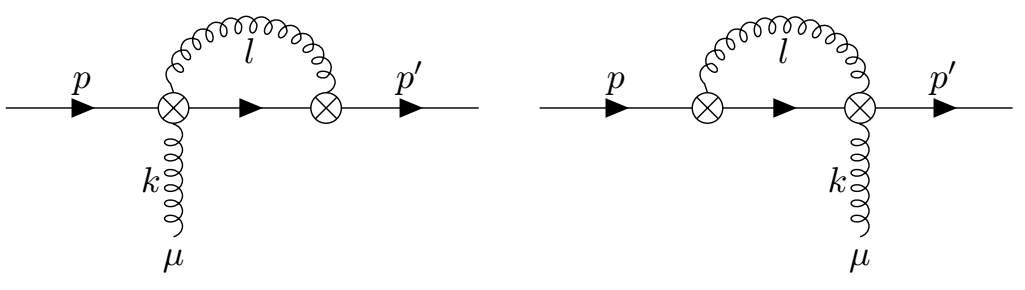}
    \caption{Spin-spin abelian interaction (sp-sp).} \label{Gamma_spsp}
\end{subfigure}
\begin{subfigure}[b]{0.45\textwidth}
    \includegraphics[scale=0.3]{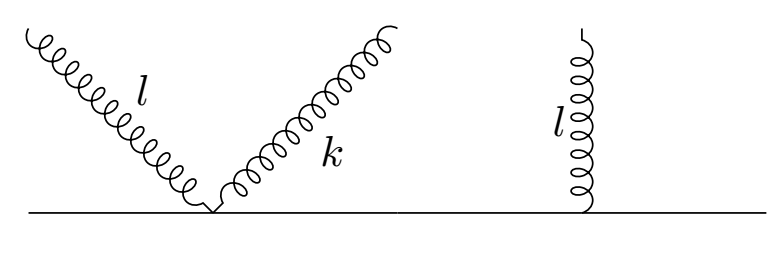}
    \caption{Left-hand ordering for the four-point vertices.} \label{Ordering1}
\end{subfigure}
\begin{subfigure}[b]{0.45\textwidth}
    \includegraphics[scale=0.35]{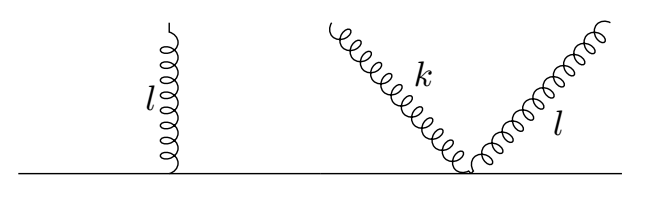}
    \caption{Right-hand ordering for the four-point vertices.} \label{Ordering2}
\end{subfigure}
\caption{ Here, from Fig.~\ref{Gamma_scscsp} to Fig.~\ref{Gamma_spsp}, we collect all second-order abelian Feynman diagrams for the on-shell quark-gluon vertex function at one-loop. In order to apply correctly the second-order rules depicted in Fig.~\ref{SecondOR_4pointvert} for the four-point vertices, we need to take into account the right order for the external gluon vertex with respect to the virtual gluon. Thus, in all second-order diagrams where the external gluon is attached to a virtual gluon, we consider the ordering as shown in Figs.~\ref{Ordering1} and \ref{Ordering2}, i. e., the external gluon is understood to be between the two lines of the virtual gluons.   }
\label{SO_diagrams_QED_vertex}
\end{figure}

\subsection{On-shell non-abelian graph}

In this section we calculate the  non-abelian vertex function at the one-loop level for the on-shell configuration of momenta, whose first-order Feynman diagram is displayed in Fig.~\ref{NonAbelian_graph} in the complete second-order formalism. We follow the same strategy that we developed for the abelian diagram in the last section, working again in the Feynman gauge ($\xi = 0$). Therefore, using the identity given in Eq.~\eqref{iden_additional_prop}, the on-shell non-abelian vertex can be written as
\begin{eqnarray}
    && \bspshort \left[\Gamma^{(NA)}\right]^{\mu a} \spshort = C_A \bspshort \bigg\{ \frac{g^3}{4m} \int \frac{d^Dl}{i(2\pi)^D} \frac{1}{q^{\prime 2} q^2 D_{l}} 
    \left( \slashed{p}^\prime + m\right) \gamma^\nu \left(-\slashed{l}+m\right) \gamma^\rho V_{\rho\nu\mu}(q,-q^\prime,k) \bigg\} \spshort \,, \label{onshell_vertex_na_1}
\end{eqnarray}
where $V_{\alpha\beta\mu}(r,s,t)$ is the three-gluon vertex defined in Eq.~\eqref{gluon_3_vertex}, $q=p+l$ and $q^\prime=p^\prime+l$ just as in the off-shell case. Moreover, we have suppressed the color factor $T^a$ for notational simplicity.
We transform the expression in 
Eq.~\eqref{onshell_vertex_na_1} into a second-order form as follows:
\begin{eqnarray}
    \bspshort \left[\Gamma^{(NA)}\right]^{\mu a} \spshort &=& C_A \bspshort \bigg\{ \frac{g^3}{4m} \int \frac{d^Dl}{i(2\pi)^D} \frac{1}{q^{\prime 2} q^2 D_{l}} A^{\nu}_{-l,p^\prime+l} A^\rho_{p,-p-l} V_{\rho\nu\mu}(q,-q^\prime,k) \bigg\} \spshort, \nonumber \\
    &=& C_A \bspshort \bigg\{ \frac{g^3}{4m} \int \frac{d^Dl}{i(2\pi)^D} \frac{1}{q^{\prime 2} q^2 D_{l}} \left[ B^\nu_{-l,q^\prime} B^\rho_{p,-q} - D_l \left( \sigma^{\nu\rho} + g^{\nu\rho} \right) \right]V_{\rho\nu\mu}(q,-q^\prime,k) \bigg\} \spshort \,,
\end{eqnarray}

\begin{figure}[h!]
\begin{subfigure}[b]{0.45\textwidth}
\includegraphics[scale=0.34]{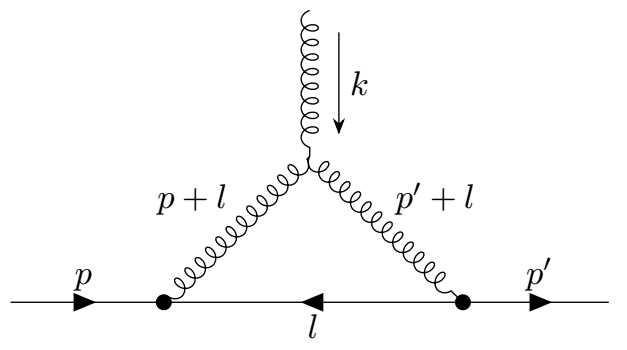}
\caption{Scalar-scalar non-abelian interaction (sc-sc).} \label{NAGamma_scsc}
\end{subfigure} 
\begin{subfigure}[b]{0.45\textwidth}
\includegraphics[scale=0.43]{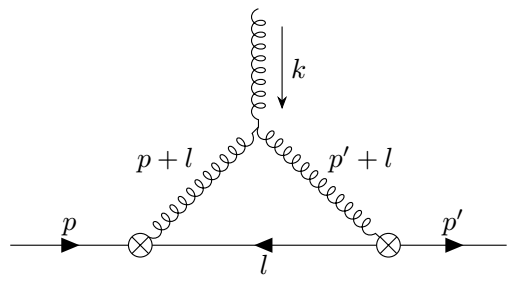}
\caption{Spin-spin non-abelian interaction (sp-sp).} \label{NAGamma_spsp}
\end{subfigure}
\begin{subfigure}[b]{\textwidth}
\includegraphics[scale=0.45]{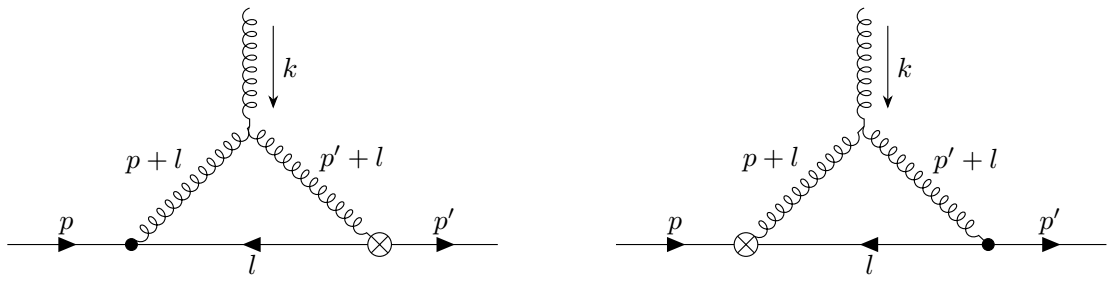}
\caption{Scalar-spin non-abelian interaction (sc-sp).} \label{NAGamma_scsp}
\end{subfigure}
\begin{subfigure}[b]{0.45\textwidth}
\includegraphics[scale=0.45]{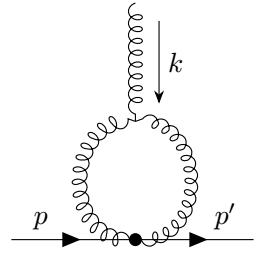}
\caption{Scalar non-abelian interaction (sc).} \label{NAGamma_sc}
\end{subfigure}
\begin{subfigure}[b]{0.45\textwidth}
\includegraphics[scale=0.45]{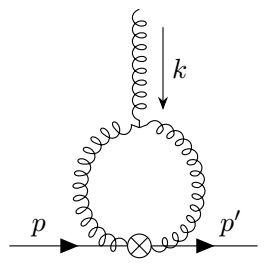}
\caption{Spin non-abelian interaction (sp).} \label{NAGamma_sp}
\end{subfigure}
\caption{ Here, from Fig.~\ref{NAGamma_scsc} to Fig.~\ref{NAGamma_sp}, we collect all second-order non-abelian Feynman diagrams for the on-shell quark-gluon vertex function at one-loop.  }
\label{SO_diagrams_NA_vertex}
\end{figure}

\end{widetext}
which can be rewritten as,
\begin{eqnarray}
    \bspshort \left[\Gamma^{(NA)}\right]^{\mu a} \spshort &=& \frac{gC_A}{4m} \bspshort \bigg( \left[\Gamma^{NA}_{\text{sc-sc}}\right]^{\mu} + \left[\Gamma^{NA}_{\text{sc-sp}}\right]^{\mu} \nonumber \\
    && \hspace*{-60pt} + \left[\Gamma^{NA}_{\text{sp-sp}}\right]^{\mu} + \left[\Gamma^{NA}_{\text{sc}}\right]^{\mu} + \left[\Gamma^{NA}_{\text{sp}}\right]^{\mu} \bigg) \spshort \,. \label{onshell_nonabelian_vertex}
\end{eqnarray}
The non-abelian vertex functions $\left[\Gamma^{NA}_{\text{a}}\right]^{\mu}$, listed in the appendix \ref{SO_vert_fun}, denote the second-order type interaction between the virtual gluons and the fermion line. For example, $\left[\Gamma^{NA}_{\text{sc-sp}}\right]^{\mu}$ denotes a scalar and spin interaction between the gluons and the fermion line, while $\left[\Gamma^{NA}_{\text{sc}}\right]^{\mu}$ denotes only a scalar type of interaction. The corresponding second-order Feynman diagrams from which the second-order vertex functions $\left[\Gamma^{NA}_{\text{a}}\right]^{\mu}$ arise, are displayed in 
Appendix~\ref{second_order_diagrams} in Fig.~\ref{SO_diagrams_NA_vertex}. 

Now, as in the previous section, the on-shell one-loop non-abelian vertex function $\left[\Gamma^{(NA)}\right]^{\mu a}$ decomposes as:
\begin{eqnarray}
    && \hspace{-4mm} \bspshort \left[\Gamma^{(NA)}\right]^{\mu a} \spshort \nonumber \\
    && \hspace{0pt} = g \bspshort \left[ \dfrac{F^{NA}_1(k^2)}{2m} (p+p^\prime)^\mu - \dfrac{F^{NA}_{\text{sp}}(k^2)}{2m} \sigma^{\mu\nu} k_\nu \right], \label{NA_onshell_vertex}
\end{eqnarray}
from which we define the form factor $F_2^{NA}(k^2)$ as,
\begin{eqnarray}
    F_2^{NA}(k^2) = F^{NA}_{\text{sp}}(k^2) - F^{NA}_{1}(k^2) \,.
\end{eqnarray}
If we follow the framework of QED, we could define an anomalous chromomagnetic dipole moment in $D=4$ dimensions by the evaluation of $F_2^{NA}(0)$. In order to compute this value, we summarize different second-order contributions to $F^{NA}_{\text{sp}}(k^2)$ as in the abelian case in Table~\ref{NA_SO_contributions_to_Fspin}. According to the result of the Appendices~\ref{SO_vert_fun} and~\ref{scalar_int_onshell} in $D=4$ dimensions, 
\begin{eqnarray}
    F_1^{NA}(0) = \dfrac{g^2 C_A}{2(4\pi)^2} \left[ \dfrac{3}{\epsilon}+4+3\ln\left(\dfrac{\overline{\mu}^2}{m^2}\right) \right]. \label{NA_F1_k0}
\end{eqnarray}

\begin{table}[h!]
\caption{\label{NA_SO_contributions_to_Fspin} \justifying{
Contributions to $F^{NA}_{\text{sp}}(0)$ that arise from the second-order non-abelian diagrams displayed in the appendix \ref{second_order_diagrams} in Fig.~\ref{SO_diagrams_NA_vertex}, grouped according to each type of second-order interactions. Here, the factor $G$ is defined as $G= \frac{g^2}{(4\pi)^2}$  }
}
\begin{ruledtabular}
\begin{tabular}{l l }
\textrm{Second-order interactions} & $ \tilde{F}_{\text{sp}}^a(0)/G$  \\
\colrule
Scalar-scalar& $ 0 $\\
Scalar-spin& $\dfrac{3}{\epsilon}+6+\ln\left(\dfrac{\overline{\mu}^6}{m_g^2 m^4}\right)$\\
Spin-spin& $\dfrac{2}{C_A} F_1^{NA}(0)$ \\
Scalar& $0$ \\
Spin& $-3\left[\dfrac{1}{\epsilon}+\ln\left(\dfrac{\overline{\mu}^2}{m_g^2}\right)\right]$
\end{tabular}
\end{ruledtabular}
\end{table}

From Table~\ref{NA_SO_contributions_to_Fspin}, we observe that, as expected, the pure scalar contributions given in the first and fourth row do not contribute to $F_{\text{sp}}^{NA}(0)$. We also observe that the second-order vertex function $\left[\Gamma^{NA}_{\text{sp-sp}}\right]^\mu$, whose corresponding Feynman diagram is depicted in 
Fig.~\ref{NAGamma_spsp}, is proportional to $F_1^{NA}(0)$. Thus, this diagram does not contribute to $F_2^{NA}(0)$. A similar result also holds true with the abelian contributions. In that case, the diagram that does not contribute to $F_2^{A}(0)$ is depicted in Fig.~\ref{Gamma_scscsp}.

Finally, from rows two and five of Table~\ref{NA_SO_contributions_to_Fspin}, we obtain
\begin{eqnarray}
    F_2^{NA}(0) = \dfrac{C_A g^2}{(4\pi)^2}\left[ 3 - \ln\left( \dfrac{m^2}{m_g^2} \right) \right], \label{F2_NA_k0}
\end{eqnarray}
which agrees with the result presented in Ref.~\cite{dipole_moments} once we set $C_A=3$. Since the result in Eq.~\eqref{F2_NA_k0} contains an infrared divergence, the chromomagnetic dipole moment defined by the evaluation of $F_2^{NA}(0)$ is not physically realizable. Thus, to define a physical chromomagnetic dipole moment from $F_2^{NA}(k^2)$, other kinematic limits for $k^2$ must be considered as shown in the Refs.~\cite{dipole_moments, dipole_moments_2}.  

\section{Conclusions} \label{Conclusions}

The second-order formalism of quantum chromodynamics (QCD) is built upon a Gordon-type decomposition applied to the product of a fermion propagator and a first-order vertex, as illustrated in Eq.~\eqref{PropVertFO}. This decomposition allows us to decouple the spin and scalar degrees of freedom in the chromodynamic interaction for a color-charged fermion. By employing this approach, we can enhance our understanding and intuition of quantum processes and functions that involve both fermions and gluons, such as the quark-gluon vertex $\Gamma^\mu(p,p^\prime,k)^{a,\mu}$.

In this article, we apply the second-order formalism to examine the off-shell and on-shell one-loop quark-gluon vertex in detail. Notice that in the off-shell case, a full implementation of the second-order formalism is not possible due to a lack of sufficient first-order vertices with which to pair the fermion propagator. Especifically, we have one less fermion propagator, preventing a complete pairing. Thus, in this case, the Gordon-type decomposition of Eq.~\eqref{PropVertFO} produces a hybrid framework that combines both first- and second-order formalisms. This hybrid approach naturally separates the longitudinal and transverse components of the vertex when the external gluon is attached directly to the fermion line, as shown in the abelian first-order diagram in Fig.~\ref{Abelian_graph}. The longitudinal and transverse components, as derived in Eq.~\eqref{LongTransAbelianVert}, are determined prior to any attempt at performing the tensor Feynman integral. By applying the tensor reduction identities presented in Eqs.~\eqref{TI_1} and~\eqref{TI_2}, we obtain relatively compact expressions for the longitudinal and transverse components, as summarized in Eqs.~\eqref{abelianl} and~\eqref{abeliant}. In contrast, for the non-abelian graph shown in Fig.~\ref{NonAbelian_graph}, the combination of first- and second-order formalisms does not lead to as straightforward a separation of the longitudinal and transverse components, since the external gluon is not attached to the fermion line. Nonetheless, even in this case, the longitudinal and transverse components, as expressed in Eq.~\eqref{NA_LT_1}, can still be extracted prior to the application of the tensor reduction formulae to the tensor integrals.

In the on-shell case, the second-order formalism can be fully implemented by inserting an additional fermion propagator into the vertex function, utilizing the identity in Eq.~\eqref{iden_additional_prop}. Within this framework, a detailed analysis of various scalar and spin interactions between the fermion line and the gluons can be conducted. This analysis is presented in section~\ref{Onshell_section}, where Tables~\ref{g_contributions} and~\ref{NA_SO_contributions_to_Fspin} summarize the contributions of different scalar and spin interactions to the form factors $F_{\text{spin}}^{NA}(0)$ and $F_{\text{spin}}^{A}(0)$, which are associated with the electromagnetic and chromomagnetic dipole moments, respectively. The Tables indicate that, at the one-loop order, only a single second-order diagram is required to reproduce the scalar form factors $F_1^{A}(0)$ and $F_1^{NA}(0)$. Specifically, for $F_1^{A}(0)$, the corresponding diagram is depicted in Fig.~\ref{Gamma_scscsp}, while for $F_1^{NA}(0)$, the relevant diagram is depicted in Fig.~\ref{NAGamma_spsp}. 

Another conclusion that can be drawn from Tables~\ref{g_contributions} and~\ref{NA_SO_contributions_to_Fspin} is that at $k^2=0$, only those diagrams containing a single spin interaction and at least one scalar interaction—excluding the diagram associated with the form factors $F_1(0)$- are relevant for the physical contributions to the dipole moments. In the non-abelian case, this corresponds to the diagrams where the virtual gluon interacts with the fermion line, exchanging spin only once, as shown in Fig.~\ref{NAGamma_scsp}. In the abelian case, the second-order diagrams relevant to the magnetic dipole moment are represented in Figs.~\ref{Gamma_spscsc},~\ref{Gamma_scsp} and~\ref{Gamma_spsc}.

In the diagrams of Fig.~\ref{Gamma_spscsc}, the external gluon exchanges momentum with the fermion line without altering the spin of the virtual fermions. The spin change of the fermion line arises from the loop of virtual gluons. These interactions reduce the value of the anomalous magnetic dipole moment by 1, as indicated in the fifth row of Table~\ref{g_contributions}. The other significant diagrams, shown in Figs.~\ref{Gamma_scsp} and~\ref{Gamma_spsc}, demonstrate that the spin of the fermion line is altered either through a spin interaction with the virtual gluons or via a spin interaction involving both the external and virtual gluons. In the former case, depicted in Fig.~\ref{Gamma_spsc}, the anomalous magnetic moment is reduced by half, while in the latter case, shown in Fig.~\ref{Gamma_scsp}, the value is increased by $\frac{7}{2}$, according to Table~\ref{g_contributions}, considering only the finite contributions. Summing these contributions yields a total value of 2, which, when multiplied by the global factor $\frac{\alpha}{4 \pi}$, results in the Schwinger value for the anomalous magnetic dipole moment of the electron. Assuming that this result generalizes to higher-order loops, we conjecture that only a specific set of second-order diagrams contributes to the anomalous dipole moments, in which the spin interchange is minimal. 

\begin{acknowledgements}

 A.B. wishes to acknowledge the {\em Coordinaci\'on de la Investigaci\'on Cient\'ifica} of the{\em Universidad Michoacana de San Nicol\'as de Hidalgo}, Morelia, Mexico,  grant no. 4.10, 
 {\em Secretaría de Ciencia, Humanidades, Tecnología e Inovación} 
 SECIHTI (Mexico), project CBF2023-2024-3544
as well as the Beatriz-Galindo support during his scientific stay at the University of Huelva, Huelva, Spain. 
V.M.B.G. also acknowledges the support of
SECIHTI (Mexico).

\end{acknowledgements}

%%%%%%%%%%%%%%%%%%%%%%%%%%%%%%%%%%%%%%%%%%%%%%%%%%%%%%%%%%%%%%%%%%%%%%%%%%%
%%%%%%%%%%%%%%%%%%%%%%%%%%%%%%%%%%%%%%%%%%%%%%%%%%%%%%%%%%%%%%%%%%%%%%%%%%%

\appendix 

\begin{widetext}
\section{Coefficients $\alpha_i$ and $\beta_i$ for the off-shell non-abelian diagram} \label{coefficients}
 
We here collect the coefficients $\alpha_i$ and $\beta_i$ in Eq.~\eqref{Vertex_NA} in terms of scalar integrals with shifted dimensions and higher powers in the propagators. 
The coefficients $\alpha_i$ are:
\begin{eqnarray} 
\alpha_1 &=& (2-D) J_1^{D+2}(1,1,1)-\xi m^2 J_1^D(2,0,1)+ \left(4 p \cdot p^\prime-\frac{m^2\xi }{2}-2 p^2\right)J_1^{D+2}(1,2,1) +2 J_1^D(1,1,0) \nonumber \\
   && +m^2 \xi  J_1^D(2,1,1) \left(p^2-2p \cdot p^\prime+p^{\prime 2}\right)  +J_1^D(1,1,1) \left(2 m^2+\xi\left[p^2-2 p \cdot p^\prime+p^{\prime 2}\right)\right] \nonumber \\
   && +J_1^{D+2}(2,1,1) \left(-\frac{m^2 \xi }{2}-2 p \cdot p^\prime+4 p^{\prime 2}\right)-\xi  J_1^D(0,1,1)-\xi  J_1^D(2,0,0) -\frac{1}{2} \xi  J_1^{D+2}(1,2,0) \nonumber \\
   && -\frac{1}{2} \xi  J_1^{D+2}(2,1,0)+\xi  p^2 J_1^D(0,2,1)+2
   \xi  p^2 J_1^{D+2}(0,3,1)-\xi  p^2 J_1^D(1,2,1) \left(p^2-2 p \cdot p^\prime +p^{\prime 2}\right) \nonumber \\
   && +\xi  J_1^D(2,1,0) \left(p^2-2 p \cdot p^\prime +p^{\prime 2}\right)-2 \xi  p^2 J_1^{D+2}(1,3,1) \left(p^2-2
   p \cdot p^\prime+p^{\prime 2}\right) -2 \xi  p^{\prime 2} J_1^{D+2}(3,0,1) \nonumber \\
   && +2 \xi  p \cdot p^\prime
   J_1^{D+2}(2,2,1) \left(-p^2+2 p \cdot p^\prime-p^{\prime 2}\right)  +2 \xi  p^{\prime 2} J_1^{D+2}(3,1,1) \left(p^2-2 p \cdot p^\prime +p^{\prime 2}\right) \,, \\
   \alpha_2 &=& m (D-\xi -1) J_1^D(1,1,1)+2 m (D-\xi -1) J_1^{D+2}(1,2,1)-m \xi  J_1^{D+2}(2,1,1)+m \xi  p^2 J_1^D(1,2,1) \nonumber \\
   && +4 m \xi  p^2 J_1^{D+2}(1,3,1) +6 m \xi  p^2 J_1^{D+4}(1,4,1)-m \xi  J_1^{D+2}(2,2,1) \left(p^2-6 p \cdot p^\prime+3
   p^{\prime 2}\right) \nonumber \\
   && +2 m \xi  J_1^{D+4}(2,3,1) \left(p^2+2 p \cdot p^\prime-p^{\prime 2}\right)+m J_1^{D+4}(3,2,1) \left(4 \xi  p \cdot p^\prime-2 \xi  p^{\prime 2}\right) \,,  \\
   \alpha_3 &=& - \Big[ (m-D m) J_1^D(1,1,1)-2 (D-1) m J_1^{D+2}(2,1,1)-m \xi  J_1^D(2,0,1)-m \xi 
   J_1^{D+2}(1,2,1) \nonumber \\
   && -2 m \xi  p^2 J_1^{D+2}(2,2,1)-2 m \xi
    p^2 J_1^{D+4}(2,3,1)+m \xi  J_1^D(2,1,1) \left(p^2-4 p \cdot p^\prime+2
   p^{\prime 2}\right) \nonumber \\
   && +2 m \xi  J_1^{D+2}(3,1,1) \left(p^2-6 p \cdot p^\prime+3 p^{\prime 2}\right)-2 m \xi  J_1^{D+4}(3,2,1) \left(p^2+2
   p \cdot p^\prime-p^{\prime 2}\right) \nonumber \\
   && +6 m \xi  \left(p^{\prime 2}-2 p \cdot p^\prime\right) J_1^{D+4}(4,1,1) -2 m \xi  J_1^{D+2}(3,0,1) \Big] \,, \\
   \alpha_4 &=& m \xi  J_1^D(0,2,1)+2 m \xi  J_1^{D+2}(0,3,1)+\frac{1}{2} m \xi 
   J_1^{D+2}(1,2,1)+\frac{1}{2} m \xi  J_1^{D+2}(2,1,1) \nonumber \\
   && -m \xi  J_1^D(1,2,1)
   \left(p^2-2 p \cdot p^\prime+p^{\prime 2}\right) -2 m \xi 
   J_1^{D+2}(1,3,1) \left(p^2-2 p \cdot p^\prime+p^{\prime 2}\right) \nonumber \\
   && +m \xi
    J_1^{D+2}(2,2,1) \left(p^2-2 p \cdot p^\prime+p^{\prime 2}\right) \,, \\
   \alpha_5 &=& - \Big[ m \xi  J_1^D(2,0,1)+\frac{1}{2} m \xi  J_1^{D+2}(1,2,1)+\frac{1}{2} m \xi 
   J_1^{D+2}(2,1,1)+2 m \xi  J_1^{D+2}(3,0,1) \nonumber \\
   && -m \xi  J_1^D(2,1,1) \left(p^2-2
   p \cdot p^\prime+p^{\prime 2}\right) +m \xi  J_1^{D+2}(2,2,1)
   \left(p^2-2 p \cdot p^\prime+p^{\prime 2}\right) \nonumber \\
   && -2 m \xi 
   J_1^{D+2}(3,1,1) \left(p^2-2 p \cdot p^\prime+p^{\prime 2}\right) \Big] \,,  \\
   \alpha_6 &=& (D+1) J_1^{D+2}(1,2,1)+4 (D-2) J_1^{D+4}(1,3,1)+2 m^2 \xi  J_1^{D+2}(2,2,1)+6 m^2
   \xi  J_1^{D+4}(1,4,1) \nonumber \\
   && +2 \xi  \left(m^2+p^2\right) J_1^{D+2}(1,3,1)+2 \xi 
   \left(m^2+p^{\prime 2}\right) J_1^{D+4}(2,3,1)-\xi  J_1^D(1,1,1)+2 \xi 
   J_1^{D+2}(1,3,0)  \nonumber \\
   && +2 \xi  J_1^{D+2}(2,2,0) +6 \xi  J_1^{D+4}(1,4,0)+2 \xi 
   J_1^{D+4}(2,3,0)+\xi  p^2 J_1^D(1,2,1)+2 \xi  p^{\prime 2}
   J_1^{D+4}(3,2,1) \,, \\
   \alpha_7 &=& - \Big[ (-D-\xi +3) J_1^{D+2}(2,1,1)+(4-2 D) J_1^{D+4}(2,2,1)+2 \xi  J_1^{D+4}(2,3,1)
   \left(-m^2+p^2+2 p \cdot p^\prime\right) \nonumber \\
   &&-2 \xi  \left(m^2-2p \cdot p^\prime\right) J_1^{D+4}(3,2,1)+\xi  \left(p^{\prime 2}-m^2\right)
   J_1^D(2,1,1)+2 \xi  \left(p^{\prime 2}-m^2\right) J_1^{D+2}(3,1,1) \nonumber \\
   && -2 \xi 
   J_1^D(1,1,1)-\xi  J_1^D(2,1,0)+(2-2 \xi ) J_1^{D+2}(1,2,1)-2 \xi 
   J_1^{D+2}(3,1,0)-2 \xi  J_1^{D+4}(2,3,0)\nonumber\\
   && -2 \xi  J_1^{D+4}(3,2,0) +\xi  p^2
   J_1^D(1,2,1)+4 \xi  p^2 J_1^{D+2}(1,3,1)+6 \xi  p^2 J_1^{D+4}(1,4,1)\nonumber \\
   && -2 \xi 
   J_1^{D+2}(2,2,1) \left(p^2-3 p \cdot p^\prime+p^{\prime 2}\right) \Big] \,,  \\
   \alpha_8 &=& - \Big[ (-D+\xi +3) J_1^{D+2}(1,2,1)+(4-2 D) J_1^{D+4}(2,2,1)+m^2 (-\xi ) J_1^D(2,1,1)-2
   m^2 \xi  J_1^{D+4}(2,3,1) \nonumber \\
   && +\xi  \left(p^2-m^2\right) J_1^D(1,2,1)+2 \xi 
   \left(p^2-m^2\right) J_1^{D+2}(1,3,1)+\xi  J_1^{D+2}(2,2,1) \left(p^{\prime 2}-m^2+2 p^2-4
   p \cdot p^\prime\right) \nonumber \\
   && -2 \xi  \left(m^2+p^{\prime 2}\right) J_1^{D+2}(3,1,1)-2 \xi  \left(m^2+p^{\prime 2}\right)
   J_1^{D+4}(3,2,1)-\xi  J_1^D(1,1,1)-\xi  J_1^D(1,2,0)-\xi  J_1^D(2,1,0)\nonumber \\
   && -2 \xi 
   J_1^{D+2}(1,3,0)+(\xi -4) J_1^{D+2}(2,1,1)-\xi  J_1^{D+2}(2,2,0)-2 \xi 
   J_1^{D+2}(3,1,0)-2 \xi  J_1^{D+4}(2,3,0) \nonumber \\
   &&-2 \xi  J_1^{D+4}(3,2,0)-6 \xi 
   p^{\prime 2} J_1^{D+4}(4,1,1) \Big] \,,
   \end{eqnarray}
   \begin{eqnarray}
   \alpha_9 &=& (D+\xi -5) J_1^{D+2}(2,1,1)+4 (D-2) J_1^{D+4}(3,1,1)+\xi  \left(m^2-3 p^2\right)
   J_1^{D+2}(2,2,1) \nonumber \\
   && +2 \xi  J_1^{D+4}(3,2,1) \left(m^2-p^2-2 p \cdot p^\prime \right)-\xi  J_1^D(2,1,1) \left(m^2+p^2-p^{\prime 2}\right)+6 \xi 
   \left(m^2-2 p \cdot p^\prime\right) J_1^{D+4}(4,1,1) \nonumber\\
   && -\xi  J_1^D(1,1,1)+\xi 
   J_1^D(2,0,1)-\xi  J_1^D(2,1,0)+\xi  J_1^{D+2}(2,2,0)+2 \xi  J_1^{D+2}(3,0,1)+2
   \xi  J_1^{D+4}(3,2,0) \nonumber \\
   && +6 \xi  J_1^{D+4}(4,1,0)-2 \xi  p^2 J_1^{D+4}(2,3,1)-2 \xi 
   J_1^{D+2}(3,1,1) \left(p^2+2 p \cdot p^\prime-p^{\prime 2}\right) \,,  \\
   \alpha_{10} &=& - \Big[ m \xi  J_1^D(1,2,1)+4 m \xi  J_1^{D+2}(1,3,1)+2 m \xi  J_1^{D+2}(2,2,1)+6 m \xi 
   J_1^{D+4}(1,4,1)+4 m \xi  J_1^{D+4}(2,3,1) \nonumber \\
   && +2 m \xi  J_1^{D+4}(3,2,1) \Big] \,, \\
   \alpha_{11} &=& -m \xi  J_1^D(2,1,1)-2 m \xi  J_1^{D+2}(2,2,1)-4 m \xi  J_1^{D+2}(3,1,1)-2 m \xi 
   J_1^{D+4}(2,3,1)-4 m \xi  J_1^{D+4}(3,2,1) \nonumber \\
   && -6 m \xi  J_1^{D+4}(4,1,1) , \\
   \alpha_{12} &=& - \Big[ -\frac{1}{2} \xi  J_1^{D+2}(1,2,1)-\frac{1}{2} \xi  J_1^{D+2}(2,1,1)-2 \xi 
   J_1^{D+2}(2,2,1) \left(p^2-2 p \cdot p^\prime+p^{\prime 2}\right) \Big] \,, 
\end{eqnarray}
while the coefficients $\beta_i$ reads as, 
\begin{eqnarray}
    \beta_1 &=& 2 m \xi ^2 J_1^{D+2}(2,3,1) \left(m^2+p^2+2 p \cdot p^\prime\right)+m \xi ^2
   \left(m^2+p \cdot p^\prime\right) J_1^D(2,2,1) \nonumber \\
   && +2 m \xi ^2 J_1^{D+2}(3,2,1)
   \left(m^2+2 p \cdot p^\prime+p^{\prime 2}\right) +m \xi ^2
   J_1^D(2,2,0)-m \xi ^2 J_1^{D+2}(2,2,1)+2 m \xi ^2 J_1^{D+2}(2,3,0)\nonumber \\
   && +2 m \xi ^2
   J_1^{D+2}(3,2,0)+6 m \xi ^2 \left(p^2+p \cdot p^\prime\right)
   J_1^{D+4}(2,4,1)  +4 m \xi ^2 J_1^{D+4}(3,3,1) \left(p^2+2 p \cdot p^\prime+p^{\prime 2}\right)\nonumber \\
   && +6 m \xi ^2 \left(p \cdot p^\prime+p^{\prime 2}\right) J_1^{D+4}(4,2,1) \,,  \\
    \beta_2 &=& -m^2 \xi ^2 J_1^D(2,2,1)-\xi ^2 \left(3 m^2+p^2\right) J_1^{D+2}(2,3,1)-3 \xi ^2
   \left(m^2+p^2\right) J_1^{D+4}(2,4,1) \nonumber \\
   && -2 \xi ^2 J_1^{D+4}(3,3,1) \left(2
   m^2+p^2+p^{\prime 2}\right)-\xi ^2 \left(3 m^2+p^{\prime 2}\right) J_1^{D+2}(3,2,1)-3 \xi ^2 \left(m^2+p^{\prime 2}\right)
   J_1^{D+4}(4,2,1) \nonumber \\
   && -\xi ^2 J_1^D(2,2,0)+\frac{1}{2} \xi ^2
   J_1^{D+2}(2,2,1)-3 \xi ^2 J_1^{D+2}(2,3,0)-3 \xi ^2 J_1^{D+2}(3,2,0)-3 \xi ^2
   J_1^{D+4}(2,4,0) \nonumber \\
   && -4 \xi ^2 J_1^{D+4}(3,3,0) -3 \xi ^2 J_1^{D+4}(4,2,0) \,, \\
    \beta_3 &=& -\frac{1}{2} \xi ^2 \left(m^2-p^2\right) \left(p^2-p^{\prime 2}\right)
   J_1^{D+2}(2,3,1)+\frac{1}{2} \xi ^2 \left(m^2-p^{\prime 2}\right)
   \left(p^2-p^{\prime 2}\right) J_1^{D+2}(3,2,1) \nonumber \\
   && -\frac{3}{2} \xi ^2
   \left(m^2-p^2\right) \left(p^2-p^{\prime 2}\right)
   J_1^{D+4}(2,4,1)+\frac{3}{2} \xi ^2 \left(m^2-p^{\prime 2}\right)
   \left(p^2-p^{\prime 2}\right) J_1^{D+4}(4,2,1)\nonumber\\
   &&+\frac{1}{2} \xi ^2
   \left(m^2-p \cdot p^\prime\right) J_1^{D+2}(2,2,1)+\frac{1}{2} \xi ^2
   J_1^{D+2}(2,2,0)+\frac{1}{2} \xi ^2 \left(p^{\prime 2}-p^2\right)
   J_1^{D+2}(2,3,0)\nonumber \\
   && +\frac{1}{2} \xi ^2 \left(p^2-p^{\prime 2}\right)
   J_1^{D+2}(3,2,0)+\frac{3}{2} \xi ^2 \left(p^{\prime 2}-p^2\right)
   J_1^{D+4}(2,4,0)+\xi ^2 \left(p^2-p^{\prime 2}\right)^2
   J_1^{D+4}(3,3,1) \nonumber \\
   && +\frac{3}{2} \xi ^2 \left(p^2-p^{\prime 2}\right)
   J_1^{D+4}(4,2,0)+\frac{3}{2} J_1^{D+2}(1,2,1)-\frac{3}{2} J_1^{D+2}(2,1,1) \,,\\
    \beta_4 &=& -m \xi ^2 J_1^D(2,2,1)-4 m \xi ^2 J_1^{D+2}(2,3,1)-4 m \xi ^2 J_1^{D+2}(3,2,1)-6 m
   \xi ^2 J_1^{D+4}(2,4,1)\nonumber \\
   && -8 m \xi ^2 J_1^{D+4}(3,3,1)-6 m \xi ^2 J_1^{D+4}(4,2,1) \,, \\
    \beta_5 &=& \frac{1}{2} m \xi ^2 J_1^{D+2}(2,2,1) \left(p^2-2 p \cdot p^\prime +p^{\prime 2}\right)-3 m J_1^D(1,1,1) \,, \\
    \beta_6 &=& -\frac{1}{2} \xi ^2 \left(m^2-p^2\right) J_1^{D+2}(2,3,1) \left(p^2-2
   p \cdot p^\prime+p^{\prime 2}\right)+\frac{1}{2} \xi ^2
   \left(m^2-p^{\prime 2}\right) J_1^{D+2}(3,2,1) \left(p^2-2
   p \cdot p^\prime+p^{\prime 2}\right)\nonumber \\
   &&-\frac{3}{2} \xi ^2
   \left(m^2-p^2\right) J_1^{D+4}(2,4,1) \left(p^2-2 p \cdot p^\prime +p^{\prime 2}\right)+\frac{3}{2} \xi ^2 \left(m^2-p^{\prime 2}\right) J_1^{D+4}(4,2,1) \left(p^2-2 p \cdot p^\prime+p^{\prime 2}\right)\nonumber \\
   &&-\frac{1}{2} \xi ^2 J_1^{D+2}(2,3,0) \left(p^2-2 p \cdot p^\prime +p^{\prime 2}\right)+\frac{1}{2} \xi ^2 J_1^{D+2}(3,2,0) \left(p^2-2
   p \cdot p^\prime+p^{\prime 2}\right)\nonumber \\
   && -\frac{3}{2} \xi ^2
   J_1^{D+4}(2,4,0) \left(p^2-2 p \cdot p^\prime+p^{\prime 2}\right)+\xi
   ^2 \left(p^2-p^{\prime 2}\right) J_1^{D+4}(3,3,1) \left(p^2-2
   p \cdot p^\prime+p^{\prime 2}\right) \nonumber\\
   && +\frac{3}{2} \xi ^2
   J_1^{D+4}(4,2,0) \left(p^2-2 p \cdot p^\prime+p^{\prime 2}\right)-\frac{3}{2} J_1^{D+2}(1,2,1)-\frac{3}{2} J_1^{D+2}(2,1,1) \,,  \\
    \beta_7 &=& 0 \,, \\
    \beta_8 &=& -\frac{1}{2} \xi ^2 J_1^{D+2}(2,2,1) \left(p^2-2 p \cdot p^\prime+p^{\prime 2}\right)-3 J_1^{D+2}(1,2,1)-3 J_1^{D+2}(2,1,1) \,.
\end{eqnarray}

\section{Second-order vertex functions for the one-loop quark-gluon vertex on-shell} \label{SO_vert_fun}

In this appendix we list all the one-loop abelian and non-abelian second-order on-shell vertex functions defined in Eqs.~\eqref{qed_vert_SO_2} and~\eqref{onshell_nonabelian_vertex}. The corresponding Feynman diagrams for these functions are displayed in the Appendix~\ref{second_order_diagrams} in Fig.~\ref{SO_diagrams_QED_vertex} for the abelian case, and in Fig.~\ref{SO_diagrams_NA_vertex} for the non-abelian case. The Feynman rules for the second-order vertex interactions are displayed in Figs.~\ref{SecondOR_3pointvert} and~\ref{SecondOR_4pointvert}.

According to Eq.~\eqref{qed_vert_SO}, the vertex functions $\left[\Gamma^A_{a}\right]^\mu$, whose corresponding second-order Feynman diagrams (omitting the color factor) are displayed in the Fig.~\ref{SO_diagrams_QED_vertex}, correspond to
\begin{eqnarray}
    \left[\Gamma^A_{\text{sc-sc-sp}}\right]^\mu &=& -\Gamfun{(2p^\prime+l) \cdot (2p+l) \sigma^{\mu\nu} k_\nu} \,, \nonumber \\ \nonumber \\ \nonumber \\
    \left[\Gamma^A_{\text{sp-sp-sc}}\right]^\mu &=& - \Gamfun{ (2l+p+p^{\prime})^\mu \sigma^{\alpha \beta}  \sigma_{\alpha \nu} l_\beta l^\nu  }  \,, \nonumber \\ \nonumber \\ \nonumber \\
    \left[\Gamma^A_{\text{sc-sc-sc}}\right]^\mu &=& \Gamfun{(2p^\prime+l) \cdot (2p+l) (2l+p+p^\prime)^\mu} \,, \nonumber \\
    \nonumber \\ \nonumber \\ %
    \left[\Gamma^A_{\text{sp-sp-sp}}\right]^\mu &=&  \Gamfun{ \sigma^{\alpha \beta}  \sigma^{\mu\nu} \sigma_{\alpha \rho} l_\beta k^\nu l^\rho } \,, \nonumber \\ \nonumber \\ \nonumber \\
    \left[\Gamma^A_{\text{sp-sc-sc}}\right]^\mu &=& -2\Gamfun{(2l+p^\prime+p)^\mu \sigma^{\alpha\beta} k_\alpha l_\beta} \,, \nonumber \\ \nonumber \\
    \nonumber \\ %
    \left[\Gamma^A_{\text{sp-sc-sp}}\right]^\mu &=& \Gamfun{l_\beta k_\nu} [ \sigma^{\mu\nu} \sigma^{\alpha \beta} (2p^\prime+l)_\alpha - \sigma^{\alpha\beta} \sigma^{\mu\nu}  (2p+l)_\alpha ] \,, \nonumber \\ \nonumber \\ \nonumber \\
    \left[\Gamma^A_{\text{sc-sc}}\right]^\mu &=& -\Gamfuns \bigg[ \dfrac{(2p^\prime+l)^\mu}{D_{p^\prime+l}} + \dfrac{(2p+l)^\mu}{D_{p+l}} \bigg] \,, \nonumber \\ \nonumber \\ \nonumber \\
    \left[\Gamma^A_{\text{sc-sp}}\right]^\mu &=& - \Gamfuns \bigg[ \dfrac{ \sigma^{\mu\nu} (2p^\prime+l)_\nu}{D_{p^\prime+l}} - \dfrac{ \sigma^{\mu\nu} (2p+l)_\nu}{D_{p+l}} \bigg] \,, \nonumber \\ \nonumber \\ \nonumber \\
    \left[\Gamma^A_{\text{sp-sc}}\right]^\mu &=& - \Gamfuns \bigg[ \dfrac{ \sigma^{\mu\nu}  }{D_{p^\prime+l}} - \dfrac{ \sigma^{\mu\nu}  }{D_{p+l}} \bigg] l_\nu \,, \nonumber \\ \nonumber \\ \nonumber \\
    \left[\Gamma^A_{\text{sp-sp}}\right]^\mu &=& - \Gamfuns \bigg[ \dfrac{ \sigma_{\nu\alpha} \sigma^{\alpha \mu}  }{D_{p^\prime+l}} + \dfrac{ \sigma^{\mu\alpha} \sigma_{\alpha \nu}  }{D_{p+l}} \bigg] l^\nu \,.
\end{eqnarray}
Rewriting the tensor integrals above in terms of scalar integrals, using the identities~\eqref{TI_1} and~\eqref{TI_2}, the spinor products $\bspshort \left[\Gamma^A_{\text{a}}\right]^\mu \spshort $ can be decomposed according to,
\begin{eqnarray}
    \bspshort \left[\Gamma^A_{\text{a}}\right]^\mu \spshort = \bspshort \left[ \bar{F}_1^{a} (p+p^\prime)^\mu - \bar{F}_{\text{sp}}^a \sigma^{\mu\nu} k_\nu \right] \,,
\end{eqnarray}
where the second-order form factor functions $\bar{F}_1^{a}$ correspond to,
\begin{eqnarray}
    \bar{F}_1^{ \text{sc-sc-sp} } &=& 0 \,, \nonumber \\
    \bar{F}_1^{\text{sp-sp-sc}} &=& \frac{g^2}{(4\pi)^{\frac{D}{2}}} (D-1)\left[ J_2^{D}(1,1,0) + 2J_2^{D+2}(2,1,0) \right] \,, \nonumber \\
    \bar{F}_1^{\text{sc-sc-sc}} &=& \frac{g^2}{(4\pi)^{\frac{D}{2}}} \Big[ 2 J_2^{D+2}(2,1,0)
    +4 \left(m^2+3 p \cdot p^\prime\right) J_2^{D+2}(2,1,1)+4 \left(m^2+p \cdot p^\prime \right) \big[J_2^{D+4}(2,2,1) \nonumber \\
    && +2 J_2^{D+4}(3,1,1) \big]+4 p \cdot p^\prime
   J_2^D(1,1,1)+J_2^D(1,1,0)-2 J_2^{D+2}(1,1,1)
    \Big] \,,  \nonumber \\
    \bar{F}_1^{\text{sp-sp-sp}} &=& -\frac{2g^2}{(4\pi)^{\frac{D}{2}}} (D-4)(m^2-p \cdot p^\prime)\left[ J_2^{D+4}(2,2,1) + 2J_2^{D+4}(3,1,1) \right] \,, \nonumber \\
    \bar{F}_1^{\text{sp-sc-sc}} &=& -\frac{4g^2}{(4\pi)^{\frac{D}{2}}} (m^2-p \cdot p^\prime)\left[ 2 J_2^{D+4}(3,1,1) + J_2^{D+4}(2,2,1) + J_2^{D+2}(2,1,1) \right] \,, \nonumber \\
    \bar{F}_1^{\text{sp-sc-sp}} &=& \frac{4g^2}{(4\pi)^{\frac{D}{2}}} (m^2-p \cdot p^\prime) J_2^{D+2}(2,1,1) \,, \nonumber \\
    \bar{F}_1^{\text{sc-sc}} &=& \frac{g^2}{(4\pi)^{\frac{D}{2}}} \Big[ -2 \left(m^2+p\cdot p^\prime\right) \left[2 J_2^{D+2}(2,1,1)+J_2^{D+4}(2,2,1)+2 J_2^{D+4}(3,1,1)\right] \nonumber \\
   && -2 J_2^D(1,1,0)+J_2^{D+2}(1,1,1)-2 J_2^{D+2}(2,1,0) \Big] \,, \nonumber \\
    \bar{F}_1^{\text{sc-sp}} &=& 0 \,, \nonumber \\
    \bar{F}_1^{\text{sp-sc}} &=& 0 \,, \nonumber \\
    \bar{F}_1^{\text{sp-sp}} &=& \frac{g^2}{(4\pi)^{\frac{D}{2}}}  (D-1) \Big[ J_2^{D+2}(1,1,1) -2(m^2+p \cdot p^\prime)\left[J_2^{D+4}(2,2,1)+2J_2^{D+4}(3,1,1)\right]  \nonumber \\
    && + 2 J_2^{D+2}(2,1,0) \Big] \,,  \label{F1_SO}
\end{eqnarray}
and the contributions to $F^A_{\text{sp}}$ read as
\begin{eqnarray}
    \bar{F}_{\text{sp}}^{\text{sc-sc-sp}} &=& \frac{g^2}{(4\pi)^{\frac{D}{2}}} \left[ J_2^{D}(1,1,0) +4(m^2+p\cdot p^\prime)J_2^{D+2}(2,1,1) +4 p \cdot p^\prime J_2^{D}(1,1,1) \right] \,, \nonumber \\
    \bar{F}_{\text{sp}}^{\text{sp-sp-sc}} &=& 0 \,, \nonumber \\
    \bar{F}_{\text{sp}}^{\text{sc-sc-sc}} &=& 0 \,, \nonumber \\
    \bar{F}_{\text{sp}}^{\text{sp-sp-sp}} &=& \frac{g^2}{(4\pi)^{\frac{D}{2}}} \Big[ (D-5)J_2^{D}(1,1,0)+2(D-4)\big[ (m^2-p \cdot p^\prime)\left(J_2^{D+4}(2,2,1) - 2 J_2^{D+4}(3,1,1) \right) \nonumber \\
    && + J_2^{D+2}(1,1,1)\big] \Big] \,,\nonumber \\
    \bar{F}_{\text{sp}}^{\text{sp-sc-sc}} &=&  \frac{2g^2}{(4\pi)^{\frac{D}{2}}}  J_2^{D+2}(1,1,1) \,, \nonumber \\
    \bar{F}_{\text{sp}}^{\text{sp-sc-sp}} &=& 0 \, , \nonumber \\
    \bar{F}_{\text{sp}}^{\text{sc-sc}} &=& 0 \, , \nonumber \\
    \bar{F}_{\text{sp}}^{\text{sc-sp}} &=& \frac{g^2}{(4\pi)^{\frac{D}{2}}} \Big[ 2(m^2-p \cdot p^\prime)\left[  J_2^{D+4}(2,2,1) - 2 J_2^{D+4}(3,1,1) \right] + 4 (m^2+p \cdot p^\prime) J_2^{D+2}(2,1,1)  \nonumber \\
    && + 2 J_2^{D}(1,1,0) + J_2^{D+2}(1,1,1) \Big] \,, \nonumber \\
    \bar{F}_{\text{sp}}^{\text{sp-sc}} &=& \frac{g^2}{(4\pi)^{\frac{D}{2}}}\Big[ 2(m^2-p \cdot p^\prime)\left[ J_2^{D+4}(2,2,1) -2 J_2^{D+4}(3,1,1)\right] + J_2^{D+2}(1,1,1) \Big] \,, \nonumber \\
    \bar{F}_{\text{sp}}^{\text{sp-sp}} &=& -\frac{g^2}{(4\pi)^{\frac{D}{2}}} (D-2)\Big[ 2(m^2-p \cdot p^\prime)\left[J_2^{D+4}(2,2,1)-2J_2^{D+4}(3,1,1)\right] + J_2^{D+2}(1,1,1)  \Big] \,.  \label{Fmag_SO}
\end{eqnarray}
To obtain the expressions in  Eqs.~\eqref{F1_SO} and~\eqref{Fmag_SO}, we use the on-shell symmetry of the scalar integrals $J_2^D(a,b,c)=J_2^D(b,a,c)$, and apply the well-known Gordon identity,
\begin{eqnarray}
    \bspshort \left[ (p+p^\prime)^\mu - \sigma^{\mu\nu} k_\nu \right] \spshort = 2m \bspshort \gamma^\mu \spshort \,.
\end{eqnarray}
Similarly, the second-order vertex functions $\left[\Gamma^{NA}_{a}\right]^\mu$ that arise in Eq.~\eqref{onshell_nonabelian_vertex}, with the corresponding second-order Feynman diagrams in Fig.~\ref{SO_diagrams_NA_vertex}, read as
\begin{eqnarray}
    \left[\Gamma^{NA}_{\text{sc-sc}}\right]^\mu &=& \GamfunNA{(p^\prime-l)^\nu (p-l)^\rho } V_{\rho \nu \mu}(q,-q^\prime,k) \,, \nonumber \\ \nonumber \\
    \left[\Gamma^{NA}_{\text{sc-sp}}\right]^\mu &=& \GamfunNA{(p^\prime-l)^\nu \sigma^{\rho \beta} q_\beta - (p-l)^\rho \sigma^{\nu \beta} q^{\prime}_\beta} V_{\rho \nu \mu}(q,-q^\prime,k) \,, \nonumber \\ \nonumber \\
    \left[\Gamma^{NA}_{\text{sc-sp}}\right]^\mu &=& -\GamfunNA{\sigma^{\nu\alpha}\, \sigma^{\rho \beta}\, q^\prime_\alpha \, q_\beta} V_{\rho \nu \mu}(q,-q^\prime,k) \,, \nonumber \\ \nonumber \\
    \left[\Gamma^{NA}_{\text{sc}}\right]^\mu &=& -\GamfunNA{g^{\nu\rho}} V_{\rho \nu \mu}(q,-q^\prime,k) \,, \nonumber \\ \nonumber \\
\left[\Gamma^{NA}_{\text{sp}}\right]^\mu &=& -\GamfunNA{\sigma^{\nu\rho}} V_{\rho \nu \mu}(q,-q^\prime,k) \,. \label{onshell_so_na_functions}
\end{eqnarray}
Using the tensor reduction formulae, Eqs.~\eqref{TI_1} and~\eqref{TI_2}, in the tensor integrals above, the spinor products $\bspshort \left[\Gamma^{NA}_{\text{a}}\right]^\mu \spshort $ can be decomposed according to,
\begin{eqnarray}
    \bspshort \left[\Gamma^{NA}_{\text{a}}\right]^\mu \spshort = \bspshort \left[ \tilde{F}_1^{a} (p+p^\prime)^\mu - \tilde{F}_{\text{sp}}^a \sigma^{\mu\nu} k_\nu \right],
\end{eqnarray}
where the second-order form factor functions $\tilde{F}_1^{a}$ correspond to,
\begin{eqnarray}
    \tilde{F}_1^{\text{sc-sc}} &=& \left(5 m^2-p \cdot p^\prime\right) J_1^D(1,1,1)+2 \left(m^2+3 p \cdot p^\prime\right) J^{D+2}_1(2,1,1)-J^{D}_1(1,0,1)+3 J^D_1(1,1,0)-J^{D+2}_1(2,0,1) \nonumber \\
    && +2 J^{D+2}_1(2,1,0) \,, \nonumber \\
    \nonumber \\
   \tilde{F}_1^{\text{sc-sp}} &=& 2 \big\{4 \left(m^2-p \cdot p^\prime\right) J^{D+2}_1(2,1,1)-2 m^2 J^D_1(1,1,1)+4 m^2
   \left[J^{D+4}_1(2,2,1)+2 J^{D+4}_1(3,1,1) \right] \nonumber \\
   && +J^D_1(1,0,1)-J^D_1(1,1,0)-J^{D+2}_1(1,1,1)-J^{D+2}_1(2,0,1)+2
   J^{D+2}_1(2,1,0)\big\} \,, \nonumber \\ \nonumber \\
   \tilde{F}_1^{\text{sp-sp}} &=& -J^D_1(1,1,1) \left[(7-2 D) m^2+p \cdot p^\prime\right]+2 J^{D+2}_1(2,1,1) \left[(4
   D-17) m^2+p \cdot p^\prime\right] +(D-4) J^D_1(1,1,0) \nonumber \\
   && +4 (D-4) m^2 \left[J^{D+4}_1(2,2,1)+2
   J^{D+4}_1(3,1,1)\right]-(D-4) J^{D+2}_1(1,1,1)+2 (D-4) J^{D+2}_1(2,1,0) \nonumber \\
   &&+2 \left(p \cdot p^\prime-m^2\right) J^D_1(1,1,1)+3 J^D_1(1,0,1)+3 J^{D+2}_1(2,0,1) \,, \nonumber \\ \nonumber \\
   \tilde{F}_1^{\text{sc}} &=& -(D-1) \left[J^{D}_1(1,1,0)+2 J^{D+2}_1(2,1,0) \right] \,, \nonumber \\ \nonumber \\
   \tilde{F}_1^{\text{sp}} &=& 0 \,, 
\end{eqnarray}
while the functions $\tilde{F}_{\text{sp}}^{a}$ are given by
\begin{eqnarray}
    \tilde{F}^{\text{sc-sc}}_{\text{sp}} &=& 0 \,, \nonumber \\ \nonumber \\
    \tilde{F}^{\text{sc-sp}}_{\text{sp}} &=& 2 \left(m^2-p \cdot p^\prime\right) J^D_1(1,1,1)+J^D_1(1,1,0)-2
   \left[J^{D+2}_1(1,1,1)+J^{D+2}_1(2,0,1) \right] \,, \nonumber \\ \nonumber \\
    \tilde{F}^{\text{sp-sp}}_{\text{sp}} &=& -(D-4) J^{D+2}_1(1,1,1)+2 \left(p \cdot p^\prime-m^2\right) J^D_1(1,1,1)+4
   J^D_1(1,0,1)+2 J^{D+2}_1(2,0,1), \nonumber \\
   \nonumber \\ %
    \tilde{F}^{\text{sc}}_{\text{sp}} &=& 0 \,, \nonumber \\ \nonumber \\
    \tilde{F}^{\text{sp}}_{\text{sp}} &=& -3 J^D_1(1,1,0) \,.
\end{eqnarray}

\section{On-shell scalar integrals at $k^2=0$} \label{scalar_int_onshell}

To evaluate the form factors $F^{A}_1(k^2)$, $F^{NA}_1(k^2)$, $F^{A}_{\text{sp}}(k^2)$ and $F^{NA}_{\text{sp}}(k^2)$ at $k^2=0$, and the contributions to the magnetic and chromomagnetic dipoles moments, we use the following identities for on-shell scalar integrals at $k^2=0$,
\begin{eqnarray}
  \nonumber \\ (4\pi\mu^2)^\epsilon\, J^{6-2\epsilon}_2(1,1,1) &=&-\left[ \dfrac{1}{2\epsilon} + \dfrac{1}{2} \ln\left(\dfrac{\bar{\mu}^2}{m^2}\right) + \dfrac{1}{2} \right] \,, \nonumber \\  \nonumber \\ 
    (4\pi\mu^2)^\epsilon\, J^{4-2\epsilon}_2(1,1,0) &=&  \dfrac{1}{\epsilon} + \ln\left(\dfrac{\bar{\mu}^2}{m^2}\right) \,, \nonumber \\ \nonumber \\ 
    (4\pi\mu^2)^\epsilon\, J^{4-2\epsilon}_1(1,0,1) &=& \dfrac{1}{\epsilon} + 2 + \ln\left( \dfrac{\overline{\mu}^2}{m^2} \right) \,, 
    \nonumber \\ \nonumber \\ 
    (4\pi\mu^2)^\epsilon\, J^{6-2\epsilon}_1(2,0,1) &=& - \dfrac{1}{2} \left[ \dfrac{1}{\epsilon} + 3 + \ln\left(\dfrac{\overline{\mu}^2}{m^2}\right) \right] \,, \nonumber \\  \nonumber \\ 
    (4\pi\mu^2)^\epsilon\, J^{6-2\epsilon}_1(1,1,1) &=& (4\pi\mu^2)^\epsilon\, J^{6-2\epsilon}_1(2,0,1) \,, \nonumber \\ \nonumber \\ 
    (4\pi\mu^2)^\epsilon\, J^{4-2\epsilon}_1(1,1,0) &=& \dfrac{1}{\epsilon} + \ln\left(\dfrac{\overline{\mu}^2}{m_g^2}\right) \,, \nonumber \\  \nonumber \\ 
    (4\pi\mu^2)^\epsilon\, J^{6-2\epsilon}_1(2,1,0) &=& - \dfrac{1}{2} \left[ \dfrac{1}{\epsilon} + \ln\left(\dfrac{\overline{\mu}^2}{m_g^2}\right) \right] \,, \nonumber \\ \nonumber \\ 
    J^{4}_2(1,1,1) &=& - \frac{1}{2m^2} \ln\left(\dfrac{m^2}{m_g^2}\right) \,, \nonumber \\
     \nonumber \\  %
    J_1^{4}(1,1,1) &=& \dfrac{1}{2m^2}\left[ -\dfrac{\pi m}{m_g} + 1 +  \ln\left(\dfrac{m^2}{m_g^2}\right) \right] \,, \nonumber \\ \nonumber \\ 
    J_1^{6}(2,1,1) &=& \dfrac{1}{4m^2} \left[ \dfrac{\pi m}{m_g} + 1 - 2 \ln\left(\dfrac{m^2}{m_g^2}\right)  \right] \,, \nonumber \\ \nonumber \\ 
    J_1^8(2,2,1) &=& -\dfrac{1}{m^2} \left[ \dfrac{\pi m}{12 m_g} + \dfrac{1}{3} + \dfrac{1}{4} \ln\left(\dfrac{m^2}{m_g^2}\right)  \right] \,, \nonumber \\ \nonumber \\ 
    J^{6}_2(2,1,1) &=& \frac{1}{2m^2} \,, \nonumber \\ \nonumber \\ 
    J^{8}_2(3,1,1) &=& - \frac{1}{12m^2} \,, \nonumber \\ \nonumber \\ 
    J^{8}_2(2,2,1) &=&  - \frac{1}{12m^2} \,.  \\ \nonumber \label{scalar_int_k0}    
\end{eqnarray}
To perform scalar integrals, we use dimensional regularization scheme in $D=4-2\epsilon$ dimensions, introducing the renormalization mass scale parameter $\mu$ and a small fictitious gluon mass $m_g$ to regularize the infrared divergences. We have also defined $\bar{\mu}^2=4\pi e^{-\gamma_E} \mu^2$, with $\gamma_E$ being the Euler's constant. Notice that all scalar functions $J_1^D(a,b,c)$ at $k^2=0$ present either an ultraviolet divergence or an infrared divergence. 
\newpage

\end{widetext}

\bibliographystyle{apsrev4-1}
\bibliography{references.bib}% Produces the bibliography via BibTeX.

\end{document}